\documentclass{aa}  
\usepackage[varg]{newtxmath}
\usepackage{newtxtext} 
\usepackage{graphicx,booktabs}
\usepackage{hyperref}                
\usepackage{natbib,twoopt}
\bibpunct{(}{)}{;}{a}{}{,} 
\hypersetup{
 	pdfauthor={Guillaume Dubus, Nicolas Guillard, Pierre-Olivier Petrucci, Pierrick Martin},
        pdftitle={Sizing up the population of gamma-ray binaries},
     	pdfkeywords={Surveys, pulsars: general, Galaxy: stellar content, Gamma rays: stars, X-rays: binaries},
        bookmarksnumbered=true,
	colorlinks=true,   
	urlcolor=blue,     
	linkcolor=blue,     
	citecolor=blue
}

\def\ls{LS~5039}
\def\hessj{HESS~J0632+057}
\def\fgl{1FGL~J1018.6-5856}
\def\lsi{LS~I~+61\degr303}
\def\psrb{PSR~B1259-63}
\def\dem{LMC P3}
\def\ph{ph\,cm$^{-2}$\,s$^{-1}$}
\def\ergs{erg\,s$^{-1}$}

\begin{document}

   \title{Sizing up the population of gamma-ray binaries}

   \author{Guillaume Dubus
		\inst{1}
		\and
	   	Nicolas Guillard
		\inst{2}
		\and
		Pierre-Olivier Petrucci
		\inst{1}
		\and
		Pierrick Martin
		\inst{3}     	
          }

   \institute{
   	Univ. Grenoble Alpes, CNRS, Institut de Plan\'etologie et d'Astrophysique de Grenoble (IPAG), F-38000, Grenoble, France
             \and
             European Southern Observatory, Karl-Schwarzschild-str. 2, D-85748 Garching, Germany
	     \and
	     Univ. Paul Sabatier, CNRS, Institut de Recherche en Astrophysique et Plan\'etologie (IRAP), F-31028, Toulouse Cedex, France
	     }

   \date{Received ; accepted ; in original form \today}

  \abstract
   {Gamma-ray binaries are thought to be composed of  a young pulsar in orbit around a massive O or Be star, with their gamma-ray emission powered by pulsar spindown. The number of such systems in our Galaxy is not known.}
   {We aim to estimate the total number of gamma-ray binaries in our Galaxy and to evaluate the prospects for new detections in the GeV and TeV energy range, taking into account that their gamma-ray emission is modulated on the orbital period.}
   {We model the population of gamma-ray binaries and evaluate the fraction of detected systems in surveys with the {\em Fermi}-LAT (GeV), HESS, HAWC and CTA (TeV) using observation-based and synthetic template lightcurves. }
   {The detected fraction depends more on the orbit-average flux than on the lightcurve shape. Our best estimate for the number of gamma-ray binaries is 101$_{-52}^{+89}$ systems. A handful of discoveries are expected by pursuing the {\em Fermi}-LAT survey. Discoveries in TeV surveys  are less likely. However, this depends on the relative amounts of power emitted in GeV and TeV domains. There could be as many as $\approx 200$ \hessj-like systems with a high ratio of TeV to GeV emission compared to other gamma-ray binaries. Statistics allow for as many as three discoveries in five years of HAWC observations and five discoveries in the first two years of the CTA Galactic Plane survey.}
   {Continued {\em Fermi}-LAT observations are favoured over ground-based TeV surveys to find new gamma-ray binaries. Gamma-ray observations are most sensitive to short orbital period systems with a high spindown pulsar power.  Radio pulsar surveys (SKA) are likely to be more efficient in detecting long orbital period systems, providing a complementary probe into the gamma-ray binary population.  }


   \maketitle
%
\section{Introduction\label{sec:intro}}
Gamma-ray binaries are systems composed of a massive star in orbit with a compact object, characterized by broad non-thermal emission peaking (in $\nu F_{\nu}$) at energies above 1 MeV. The latter feature distinguishes them from high-mass X-ray binaries (HMXBs), whose spectral energy distribution peaks in X-rays, whereas the former distinguishes them from recycled binary millisecond pulsars, which have a low-mass companion. The compact object in gamma-ray binaries is likely to be a young, rotation-powered neutron star, with the non-thermal radiation due to the interaction of energetic pulsar wind particles with the stellar wind and radiation field of the O or Be companion.  There is ample indirect evidence for this ``binary pulsar wind nebula'' scenario even though scattering in the stellar wind prevents detection of the expected radio pulsar in most gamma-ray binaries (see \citealt{2013A\string&ARv..21...64D} for a review). Accordingly, we explicitely assume in the following that the compact object in gamma-ray binaries is a pulsar. However, many of our results are equally applicable if the gamma-ray emission is  powered by non-thermal jet emission from an accreting black hole \citep[e.g.][]{2017arXiv170401335M}. Clear evidence for gamma-ray jet emission exists for the accreting sources Cyg X-1 and Cyg X-3 but they are not gamma-ray binaries according to our definition because they are orders-of-magnitude more luminous in X-rays than in gamma rays. 

There are six gamma-ray binaries detected in high (HE, 0.1-100 GeV) or very high energy (VHE, $>$100 GeV) gamma rays. Of these, two were initially detected as HE gamma-ray sources in all-sky surveys (\object{\lsi}, \citealt{1978Natur.272..704G}, and \object{\fgl}, \citealt{2012Sci...335..189F}), two were independently detected in HE gamma rays and X-rays before the association was made (\object{\ls}, \citealt{2000Sci...288.2340P}, and \object{\dem}, \citealt{2016ApJ...829..105C}), one was detected serendipitously in VHE observations of the Monoceros Loop (\object{\hessj}, \citealt{2009ApJ...690L.101H}), and one was detected in a radio pulsar survey (\object{\psrb}, \citealt{1992ApJ...387L..37J}). Follow-up observations established that these sources are binaries harbouring a massive star and that their non-thermal emission is modulated on the orbital period.  In addition to those six gamma-ray binaries there are also four systems discovered in radio surveys with a young pulsar in orbit with a massive star, but where variable gamma-ray emission associated with the binary has yet to be detected because of the low pulsar power, long orbital timescale, and/or large distance:  \object{PSR J0045-7319}, \object{PSR J1638-4725}, \object{PSR J1740-3052} and \object{PSR J2032+4127} (see \citealt{2001MNRAS.325..979S,2011MNRAS.412L..63B,2012MNRAS.425.2378M,2015MNRAS.451..581L} respectively).

Gamma-ray binaries are probably a short-lived phase in the evolution of massive star binaries, following the birth of the neutron star and preceding  the HMXB phase, when the neutron star accretes material captured from the stellar wind instead of holding it back (see \citealt{2006csxs.book..623T} for a review on the formation of compact objects in binaries). Accretion occurs if the ram pressure from accreting matter is able to overcome the pulsar wind, turning off the pulsar mechanism \citep{1971SvA....15..342S,1975A\string&A....39..185I,1994A\string&A...282...61L,1995A\string&A...297..385C}. A gamma-ray binary can thus transition to a HMXB on the typical spindown timescale of young pulsars, a few 10$^5$ years. The evolution of the companion eventually leads to a second supernova with the formation of another compact object. Therefore, besides the unique opportunities they provide to understand the physics of pulsar winds, gamma-ray binaries also offer a window into the pulsar and orbital parameters of systems that remain bound after a supernova, and constrain the formation paths to double neutron stars and coalescing compact objects.

Achieving these goals depends on our ability to explore the population of gamma-ray binaries. The number of gamma-ray binaries in our Galaxy has been estimated from a few dozen to a few thousand systems from population synthesis studies of HMXB evolution \citep{1989A\string&A...226...88M,1995ApJS..100..217I,1996A\string&A...309..179P,1998A\string&A...332..173P}. Gamma-ray binaries are more likely to stand out in gamma rays rather than in radio, optical or X-ray surveys where they are usually inconspicuous. The discovery of \dem\ in the Large Magellanic Cloud suggests that we may have already accessed most of the observable gamma-ray binary population in our own Galaxy \citep{2016ApJ...829..105C}. 

Here, we aim to provide the first detailed  estimate of the number of gamma-ray binaries based on HE and VHE observations and to evaluate the prospects for further discoveries. To do this, we simulate observations of gamma-ray binaries to assess the probability of detections in mock gamma-ray surveys, designed to follow as closely as possible those performed or planned with the {\em Fermi}-LAT, HESS, HAWC and CTA (\S\ref{sec:surveys}). One difficulty in assessing the detectability is that the gamma-ray flux can vary strongly with orbital phase. We use input gamma-ray orbital lightcurves based on templates constructed from observations (\S\ref{sec:extrapolating}) or on a radiative model (\S\ref{sec:pop}). The estimated population size and expectations for future detections are discussed in \S\ref{sec:disc}. 

\section{Simulating surveys\label{sec:surveys}}

We simulate a measurement as the flux average of the gamma-ray binary lightcurve integrated over a certain duration and energy range. The duration of the measurement, the number of measurements (visits) and their distribution throughout time, varies according to the type of instrumentation. The observability and the detectability of the system will depend on the assumptions made for each type of survey that will be simulated. The observability depends only on the part of the sky surveyed and the location of the binary system. The detectability depends on the sensitivity of the survey, the cadence of the visits, and the emission properties of the system. 

We simulate five types of surveys with properties as close as possible to existing or envisioned surveys, without carrying out a full end-to-end simulation of the observations and of the data analysis chain: in our opinion, current knowledge on the radiative mechanisms in gamma-ray binaries does not justify performing such complex and costly end-to-end simulations. The level of detail in our mock surveys is appropriate for the basic emission model that we develop in \S\ref{sec:pop}, which represents gamma-ray binary spectra at 1 GeV and 1 TeV with mono-energetic electrons. In the GeV domain, we simulate the {\em Fermi}-LAT third catalogue ("3FGL-like") and the {\em Fermi} All-sky Variability Analysis ("FAVA-like"). In the TeV domain, we simulate the HESS Galactic Plane survey ("HESS-like"), a "HAWC-like" survey, and the CTA Galactic Plane survey ("CTA-like"). A source is considered detected if its average flux in \ph\ during the observation exceeds the threshold of the survey as defined below. We do not address the issue of how the detected gamma-ray source is identified as a gamma-ray binary, presumably through multi-wavelength follow-up observations. In particular, we make no attempt to investigate how binaries can be identified through a period analysis, like the one performed on the {\em Fermi}-LAT catalog by \citet{2012Sci...335..189F}. Here, the orbital modulation only intervenes as the source of flux variability between observation windows.

\subsection{The 3FGL-like survey\label{sec:3fgl}}

This survey tests whether the binary would have been included in the third {\em Fermi}-LAT catalogue \citep{2015ApJS..218...23A}. The whole Galactic Plane is covered so the gamma-ray binary observability is 100\%. The measurement is assumed to last four years, ignoring any time variation in exposure. The threshold for detection is taken to be 10$^{-9}\,$\ph\ (1-100 GeV) based on the flux distribution of sources detected within 10\degr\ of the Galactic Plane in \citet{2015ApJS..218...23A} (see their figure 24). We set the energy threshold at 1 GeV because the high-energy component of binaries peak around this energy and because the Galactic diffuse emission, which we do not take into account, is weaker than at 100 MeV. The 3FGL catalogue includes \ls, \lsi, and \fgl. \dem\ is also part of the catalogue but is confused with other sources in the LMC. We  also consider, where indicated, the impact of continued {\em Fermi}-LAT observations in the future. This ``extended'' 3FGL survey assumes a  detection threshold  lowered by a factor 2 and an exposure increased by a factor 4 (16 years of observations).

\subsection{The FAVA-like survey}

This is based on the search for 5.5$\sigma$ deviations from a long-term average model of the GeV emission observed with the {\em Fermi}-LAT \citep{2016arXiv161203165A}. 
Following the FAVA procedure, deviations are searched for on a weekly timescale, which sets the duration of the simulated measurement, over a time span of 8 years. 
Again, any time-variation of the exposure is ignored and observability is 100\%. 
The system is considered detected by this survey if its weekly average flux minus its long-term average flux (over 8 years) is greater than $10^{-6.5}\,$\ph (>100 MeV). 
Although the exact threshold changes as a function of location in the Galactic Plane and spectrum, this choice should be conservative based on 
 Fig.~4 of \citet{2016arXiv161203165A}. 
The FAVA survey is potentially more sensitive than the 3FGL survey to systems like \psrb\ with long orbital periods and short duty cycles for GeV emission. 
The second FAVA catalogue includes \lsi\ and \psrb. 

\subsection{The HESS-like survey}

This is based on the survey of the Galactic Plane carried out by the HESS collaboration since 2004 and is meant to be representative of what current imaging arrays of Cherenkov telescopes (IACTs) can achieve. The survey covers only part of the Galactic Plane, $-110\degr\leq l \leq 65\degr$. The observability of a system is decided by checking that it is observable for at least two hours at some point in the year at a zenith angle smaller than 45\degr, assuming the geographical location of the HESS array, and that its longitude is within the surveyed area. To produce a schedule of observations, we randomly distributed 25 visits of 2 hours over a time span of 8 years i.e. we assume a uniform survey exposure of 50 hours is achieved. We take into account that observations occur at night, ensuring  each binary has a prefered observation season. However, we do not account for Moon-less operations, which influences the distribution of observable time on a monthly timescale. The latter effect averages out over a timescale of a few years, whereas the former (prefered season) does not. The measured fluxes from each visit are then averaged and compared to a detection threshold of 3.6$\times10^{-13}\,$\ph\ ($>$1 TeV). 
This threshold corresponds to a flux of 20 mCrab\footnote{For the VHE surveys, we have converted Crab units to integrated flux above 1 TeV using 1 Crab $\equiv 1.82\times10^{-11}\,$\ph, based on the Crab spectrum measured by \citet{2008ApJ...674.1037A}.}. 
The exposure times and sensitivity are comparable to those of the HESS survey\footnote{\href{https://www.mpi-hd.mpg.de/hfm/HESS/pages/home/som/2016/01}{https://www.mpi-hd.mpg.de/hfm/HESS/pages/home/som/2016/01}}. We also considered whether a detection could be claimed from a single visit, scaling the threshold by a factor $(50/2)^{1/2}$.

\subsection{The CTA-like survey}

The CTA-like survey is intended to test the potential performance of the CTA array in detecting new gamma-ray binaries. The guiding principles are identical to the HESS-like survey. We assume that the survey is divided up into two blocks carried out in parallel during the first two years of operations, based on the plans for an initial Galactic Plane survey by the CTA consortium \citep{ctasurvey}. The first block, carried out by the South array in Chile, covers longitudes $-60\degr \leq l\leq 60\degr$  down to a sensitivity of 2.7 mCrab using 6 visits of 2 hours. The second block, carried out by the North array in the Canary Islands, covers $60\degr\leq l \leq 150\degr$ down to 4.2 mCrab in 4 visits of 2 hours. We also consider the ``full" survey covering all the Galactic Plane and carried out over a timespan of 10 years (see Fig. 6 in \citealt{ctasurvey} for details). The observability of each system is decided as for the HESS-like survey using the planned locations for the arrays.

\subsection{The HAWC-like survey}

Finally, we tested for the detection of binaries using the extended air shower array HAWC. The high duty cycle and full-sky monitor capacity of HAWC can make it more sensitive to flaring gamma-ray binaries than IACTs like HESS and CTA. Here, the binary is observable if it transits with a zenith angle smaller than 45\degr\ at the location of the HAWC array in Mexico. We then simulate one measurement per day at the time of transit and with a duration equal to transit duration. The HAWC sensitivity after five years of operation is comparable to that achieved in the HESS Galactic Plane survey, 20 mCrab above 1 TeV, assuming a source transit duration across the sky of six hours \citep{2016JPhCS.761a2034C}.  The threshold for daily detection is close to 1 Crab for a 6-hour transit, corresponding to the transit time of a source that passes  close to zenith {\em i.e.} with a declination close to +19\degr. The dependence of the threshold with source declination is taken into account by using the curve showing sensitivity as a function of declination for a $E^{-2.5}$ spectrum in Fig.~10 of  \citet{2017ApJ...843...40A}. We test for detection in each daily measurement and in the accumulated exposure over 5 years of HAWC operations.

\section{Extrapolating from observed gamma-ray binaries\label{sec:extrapolating}}

We currently have five binaries with measured orbital modulations at both GeV and TeV energies and one with a GeV modulation (\dem). How sensitive are the surveys to the detection of these binaries? This is estimated here by constructing a template lightcurve for each of the known gamma-ray binaries and, after proper scaling for distance, deriving the probability for detection once the binary is randomly located in the Galaxy.

\subsection{Template lightcurves\label{sec:template}}
\begin{figure*}
\begin{center}
\includegraphics[width=0.8\linewidth]{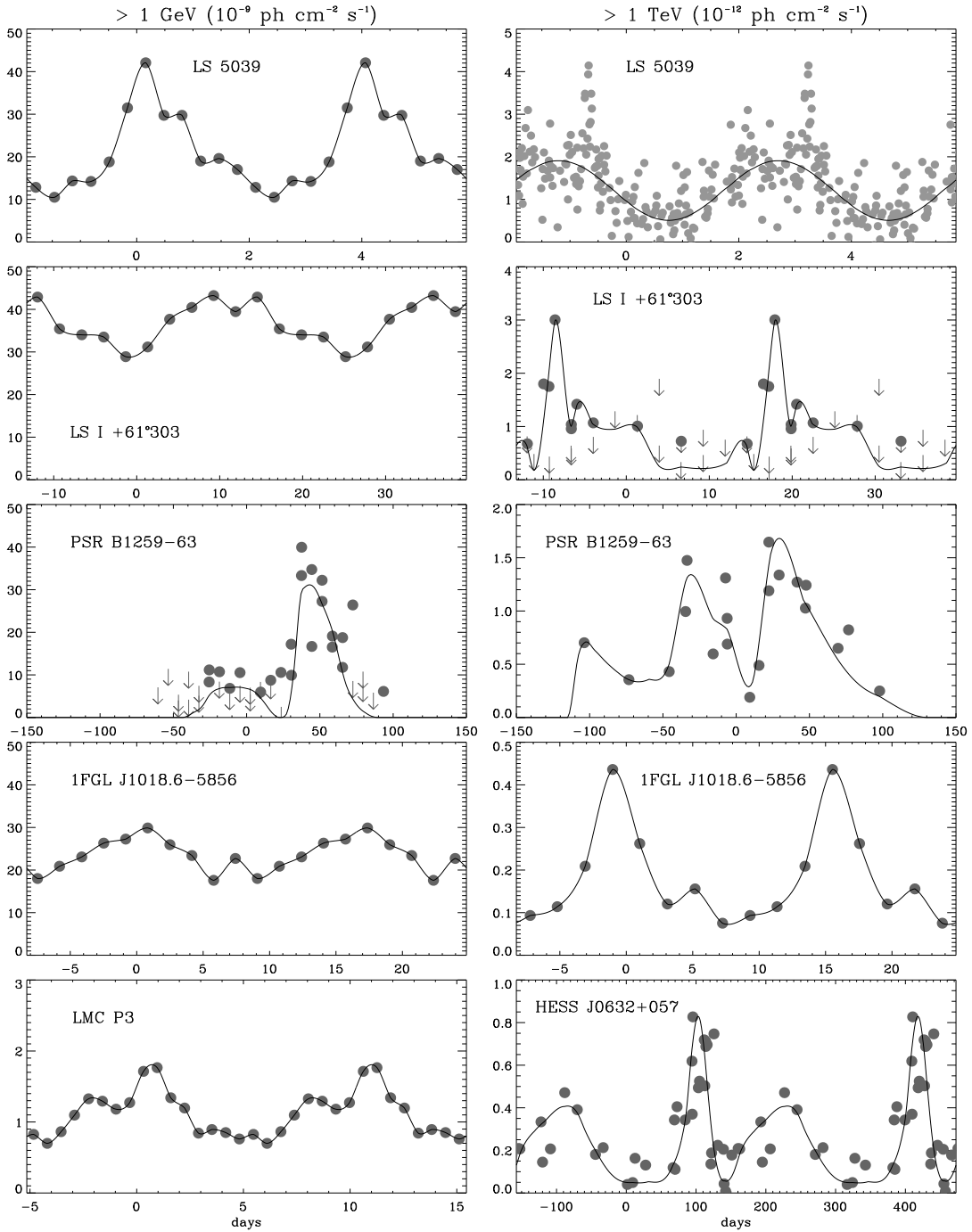} 
\caption{Template lightcurves for known gamma-ray binaries. Two orbits are shown except for \psrb\ where the plot focuses on periastron passage. The various measurements are shown in grey, with error bars omitted for clarity. Arrows indicate measurement upper limits.  Left: photon flux integrated above 1 GeV in units of $10^{-9}$\,\ph\ based on {\em Fermi}-LAT measurements. Right: photon flux integrated above 1 TeV in units of $10^{-12}$\,\ph\ based on IACT measurements. The GeV and TeV data are taken from \citet{2009ApJ...706L..56A} and \citet{2006A\string&A...460..743A} for \ls\, with $\Gamma_{\rm GeV}=2.54$; \citet{2012ApJ...749...54H} and \citet{2011ApJ...738....3A} for \lsi, with $\Gamma_{\rm GeV}=2.42$ and $\Gamma_{\rm TeV}=2.6$; \citet{2016arXiv161003264B} for \psrb, with $\Gamma_{\rm GeV}=\Gamma_{\rm TeV}=2.7$; \citet{2012Sci...335..189F}  and \citet{2015A\string&A...577A.131H} for \fgl, with $\Gamma_{\rm GeV}=3.1$ and $\Gamma_{\rm TeV}=2.7$. The GeV data for \dem\ is from \citet{2016ApJ...829..105C} with $\Gamma_{\rm GeV}=2.8$. The TeV data for \hessj\ is from \citet{2014ApJ...780..168A}.}
\label{fig:object_lc}
\end{center}
\end{figure*}

Figure~\ref{fig:object_lc} shows template lightcurves for each of the known gamma-ray binaries based on the  GeV and TeV observations available at the time of writing. In most cases, we simply took the phase-folded measurements and interpolated using splines. For \psrb, \lsi, and \hessj, the error bars, phase coverage, or orbit-to-orbit variations make it difficult to assess the mean orbital lightcurves. In these cases, our templates are meant to be representative of the behaviour of the system in that they roughly capture the amplitude and phase variations that have been observed. The template GeV and TeV lightcurves are given in \ph, integrated above 1 GeV and 1 TeV (respectively). We converted to these units assuming a simple power-law when the data was not directly available in this format. The source of the data and the power-law photon index $\Gamma$ (with $dN\propto E^{-\Gamma}dE$) that we used can be found in the caption to Fig.~\ref{fig:object_lc}. Given the low statistiscs, the GeV lightcurve of \hessj\ (not shown in Fig.~\ref{fig:object_lc}) is described as a two bin lightcurve (orbital phases 0.0-0.5 and 0.5-1.0) using the spectral parameters in Tab.~1 of \citet{2017arXiv170704280L}, who report the first detection of this system at GeV energies\footnote{The {\em Fermi}-LAT detection reported by  \citet{2016MNRAS.463.3074M}   is compatible with the detection of the low-energy end of the VHE spectrum rather than the detection of a distinct GeV spectral component as in the other gamma-ray binaries.}.

Table~\ref{tab:detected} lists the detected systems for each mock survey presented in \S\ref{sec:surveys}, given the template lightcurves and locations of the known binaries in the Galaxy. \lsi\ is not detected in the HESS-like survey due to its location. \ls\ and \lsi\ are observable with HAWC but undetected due to their unfavorable declinations, in agreement with the 18 months of HAWC observations that have been reported to date \citep{2017ApJ...843...40A}.  \psrb\ is always detected in the FAVA survey. The system is just below the threshold of the 3FGL survey when the four-year survey timespan includes only one periastron passage of the 3.4\,yr orbit, as observed. \psrb\ has a very small probability ($<1\%$) of being detected in the HESS survey due to its low duty cycle and flux, but a 50\% chance of being detected in the CTA-like survey. \hessj\ is outside the HESS and initial CTA-like surveys; it is detected in the full CTA survey. \fgl\ is detected only in the 3FGL-like survey and in the full CTA-like survey. \dem\ is detected only in the 3FGL-like survey. These results are fully consistent with the actual 3FGL, FAVA, and HESS survey observations.
\begin{table*}
\caption{Detected systems in the mock surveys based on the template lightcurves in Fig.~\ref{fig:object_lc}.}
\label{tab:detected}
\centering
\begin{tabular}{l l}
\toprule
\toprule
mock survey & detected system \\
\midrule
3FGL 	& \ls, \lsi, \fgl, \dem \\
FAVA 	& \lsi, \psrb\\
HESS 	& \ls\\
HAWC	& none \\
CTA		& \ls, \lsi, \psrb\\
CTA (full)	& \ls, \lsi, \psrb, \hessj, \fgl\\
\bottomrule
\end{tabular}
\end{table*}

\subsection{Galactic distribution\label{sec:gal_distribution}}
\begin{figure}
\begin{center}
\includegraphics[width=0.7\linewidth]{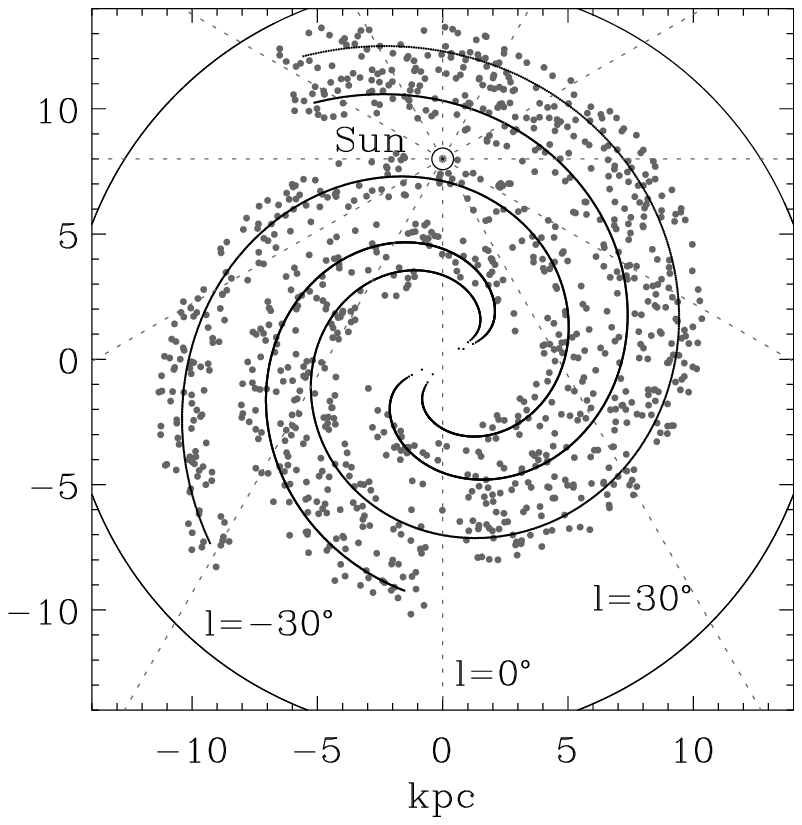} 
\includegraphics[width=0.7\linewidth]{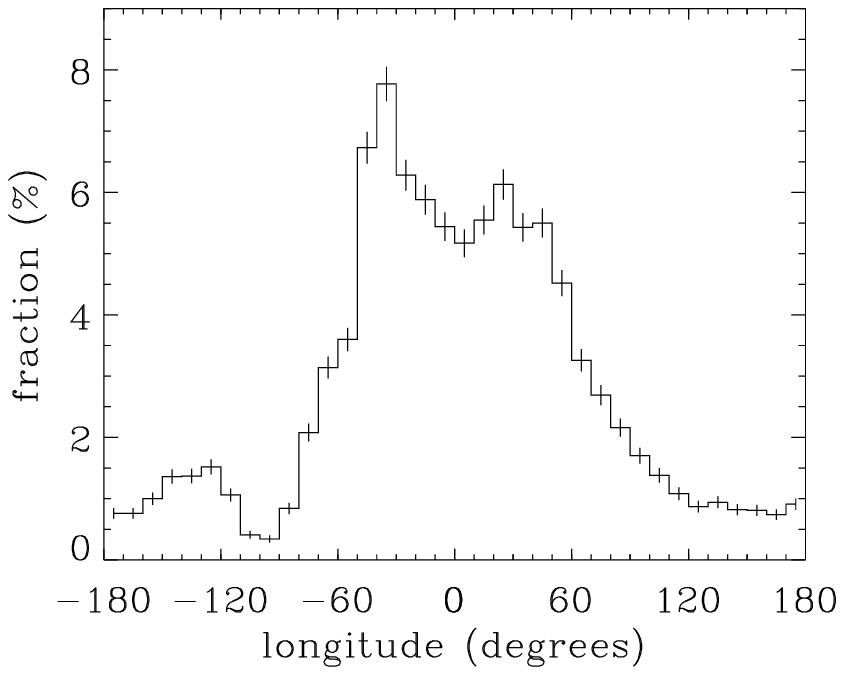} 
\caption{Top: map of randomly-generated locations for gamma-ray binaries in the Galaxy. Bottom: longitude distribution of gamma-ray binaries under the assumptions described in \S\ref{sec:gal_distribution}.}
\label{fig:gal_distribution}
\end{center}
\end{figure}

We assume that gamma-ray binaries are located in or close to the spiral arms of our Galaxy, like O and B stars and the HMXBs to which they are directly related.   The Galaxy is modelled as four one-dimension spiral arms. We use the arm formula of  \citet{2009MNRAS.397..164R} with parameters adjusted to reproduce the Galactic structure in Figure 5 of \citet{2003A\string&A...397..133R}. Our Sun is 8 kpc away from the Galactic Center. Binaries are spread out uniformly across the Galaxy disk (15 kpc), keeping only those within 1 kpc of a spiral arm and more than 3 kpc away from the Galactic Center (to account for the older stellar population in the bulge, see Fig.~\ref{fig:gal_distribution}). The binaries are assumed to reside in the Galactic Plane ($b=0\degr$). The model Galactic longitude distribution (Fig.~\ref{fig:gal_distribution}) compares well to the HMXB longitude distribution plotted in \citet{2002A\string&A...391..923G} or \citet{2015A\string&ARv..23....2W}. 

The ground-based surveys (HESS, HAWC, initial CTA) access only part of the Galactic Plane, hence only a fraction of the binaries are observable for them. These fractions are given in Tab.~\ref{tab:observable} for both a distribution along spiral arms, as described above, and a strictly uniform disk distribution. This makes little difference. In the following, we considered only the spiral arm distribution. 

\begin{table}
\caption{Fraction (\%) of observable binaries in each survey.}
\label{tab:observable}
\centering
\begin{tabular}{l c c c}
\toprule
\toprule
 & HESS & HAWC & CTA \\
\midrule
spiral & 72.25$\pm0.28$ & 55.85$\pm0.31$ & 82.94$\pm0.23$ \\
disk   & 75.26$\pm0.27$ & 52.44$\pm0.31$ & 80.94$\pm0.24$\\
\bottomrule
\end{tabular}
\end{table}

\subsection{Detectable fraction based on observed lightcurves\label{sec:lc}}

Table~\ref{table:detectable_grbi} reports the detectable fraction of gamma-ray binaries in the various surveys, based on a sample of $10^4$ systems randomly distributed in Galactic location and in starting orbital phase for each template lightcurve in Fig.~\ref{fig:object_lc} after scaling for distance. The lightcurves are then run through the simulated observations of each survey described in \S\ref{sec:surveys} to test for detection. For example, the 3FGL-like survey detects 77.9$\pm$0.8\% of the binaries that are assumed to have a GeV lightcurve identical to \ls. Here and in the remainder of the paper, the errors represent the 95\% confidence interval derived from a likelihood analysis\footnote{Here, we estimate the probability $p$ to detect a binary in a survey. We find $m$ detections after running a random sample of $n$ systems through our mock survey procedure. The likelihood function is ${L}(p)=C_n^m p^m (1-p)^{n-m}$, where $C_n^m$ is the binomial coefficient. The function has a maximum ${L}_m$ for $p=m/n$. Defining the test statistic as ${T}=2\log ({ L}_m/L)$ and applying Wilks' theorem, the 95\% confidence interval on $p$ is calculated from the interval where ${T}\leq 3.84$ (the cutoff value in a $\chi^2$ distribution with one degree of freedom beyond which the probability $\leq 5$\%). The chosen number of systems $n$ to simulate is a compromise between computational time and statistical error.}.

\fgl\ and \dem\ have high enough luminosities that such systems are detectable anywhere in the Galaxy with the 3FGL survey. Unsurprisingly, the FAVA-like survey is best-suited for low duty cycle lightcurves like \psrb. Our analysis recovers that \psrb\ is detected in FAVA, but would not include \lsi. The latter is inconsistent with the FAVA catalogue and is the result of neglecting the orbit-to-orbit variations that are detected in this source \citep{2013ApJ...773L..35A}. The FAVA-like analysis also fails to detect \ls\ and \fgl\ since the amplitude of the flux variations on weekly timescales are insufficient to flag them. In this case, this is consistent with the actual FAVA results \citep{2016arXiv161203165A}.

The numbers remain small in the HESS and HAWC-like surveys. It is only with CTA that detection fractions comparable to those achieved by the {\em Fermi}-LAT will be possible due to the high sensitivity of the instrument. Very few systems are detected as transients in the ground-based surveys ({\em i.e.} detected only in one visit): the flux increase compared to the average in a highly eccentric system rarely compensates the higher sensitivity threshold for observations on a shorter duration. Hence, the fraction of detected systems in VHE surveys decreases with longer orbital period even if the systems have comparable maximum TeV luminosities due to a high eccentricity.

The fractions in Tab.~\ref{table:detectable_grbi} give an estimated detection probability from which we can constrain the maximum size of the underlying population\footnote{We take the detection probability $p$ derived by the simulation and find the population $n$ that maximises the likelihood (see footnote 4) to detect $m$ observed systems. Here, $m=1$ for each type of gamma-ray binary.}. Any other existing system with identical properties to \fgl\ or \dem\ would have been detected since the detection probability is 1. For \ls, knowing that the system is detected in the 3FGL survey, the 77.9\% probability implies with $>$95\% confidence that there is at most three systems with comparable lightcurves in our Galaxy and most likely only one. The same conclusion is reached for \lsi. For \psrb, given the FAVA detection,  the 12.6\% detection probability implies 7$^{+26}_{-6}$ such systems in our Galaxy. Therefore, on average, there may be one more \ls\ or \lsi\ type system, and 6 other \psrb-like systems in the Galaxy that could have escaped detection in the {\em Fermi}-LAT data.

The VHE detection probabilities are not as constraining as the HE ones except for \hessj. The detection probability is only 0.8\% in both the 3FGL and HESS-like surveys. The lack of detection in those surveys (note that \hessj\ is outside the HESS survey area, \S\ref{sec:template}) places an upper limit of $<$231 on the number of \hessj-like systems in the Galaxy. The initial CTA-like survey should detect $11^{+8}_{-6}$  of those 231 systems, or will reduce their estimated number to $8^{+30}_{-7}$ should it detect only \hessj\ after the full 10 year Galactic Plane survey. CTA will thus be able to strongly constrain the number of such systems.

\begin{table*}
\caption{Fraction of detected systems in each survey using the lightcurves in Fig.~\ref{fig:object_lc} as templates (see \S\ref{sec:lc}).}
\label{table:detectable_grbi}
\centering
\begin{tabular}{lrrrrrrr}
\toprule
\toprule
				& LS\,5039			& LS\,I\,+61\degr303 	&  PSR\,B1259-63 	& HESS\,J0632+057 	& 1FGL\,J1018.6-5856	& LMC P3  \\
\midrule
P$_{\rm orb}$ (days) & 3.9				& 26.5				& 	1236.7		&  315				& 16.5				& 10.3\\
eccentricity		& 0.35				& 0.54				& 0.87			& 0.83				& $-$ 				&  $-$ 	\\
distance (kpc)		& 2.9 				& 2.0					& 2.3				& 1.6					&	5.4 				& 50 	\\
$F_{\rm max, GeV}$ (ph\,s$^{-1}$)	& 4.2$\times 10^{37}$ & 2.1$\times 10^{37}$	& 2.0$\times 10^{37}$	& 2.9$\times 10^{35}$			&	1.0$\times 10^{38}$ 		& 5.4$\times 10^{38}$ \\
$F_{\rm max, TeV}$	(ph\,s$^{-1}$)	& 1.9$\times 10^{33}$ & 1.4$\times 10^{33}$	& 1.1$\times 10^{33}$	& 2.5$\times 10^{32}$ &	1.5$\times 10^{33}$ 		& $-$ \\
\midrule
\multicolumn{5}{l}{{\em HE surveys} (\%)}\\
FAVA  			& 0.6$\pm0.2$			& 1.4$\pm0.2$ 		& 12.6$\pm0.7$ 		& $<$0.1			& 8.0$\pm0.5$			& 30.6$\pm 0.9$\\
3FGL 			& 77.9$\pm 0.8$ 		& 67.1$\pm0.9$ 		& 3.5$\pm0.4$ 		& 0.8$\pm0.2$ 	& 100				& 100\\
3FGL (extended)	& 100 				& 97.3$\pm0.3$		& 7.7$\pm0.5$ 		& 1.9$\pm0.3$ 	& 100				& 100\\
\midrule
\multicolumn{5}{l}{{\em VHE surveys} (\%)}\\
HESS 			& 10.3$\pm0.6$ 		& 3.2 $\pm 0.4$ 		& 1.5$\pm0.3$	 		& 0.8$\pm0.2$		& 5.2$\pm0.5$			& $-$\\
HAWC 			&   7.7$\pm0.5$ 		& 2.1 $\pm 0.3$ 		& 0.4$\pm0.1$ 		& 0.3$\pm0.1$		& 4.0$\pm0.4$			& $-$\\
CTA 				& 65.8$\pm0.9$ 		& 23.7$\pm0.8$ 		& 7.0$\pm0.5$ 		& 5.1$\pm0.4$		& 35.2$\pm 0.9$		& $-$\\
CTA 	(full)			& 98.0$\pm0.3$ 		& 47.0$\pm1.0$ 		& 21.2$\pm0.8$ 		& 11.2$\pm0.6$	& 70.0$\pm 0.9$		& $-$\\
\bottomrule
\end{tabular}
\end{table*}

\begin{table*}
\caption{Fraction of detected systems in each survey using synthetic lightcurves as templates (see \S\ref{sec:synth}).}
\label{table:synthetic_grbi}
\centering
\begin{tabular}{l r r r r r r r }
\toprule\toprule
				& LS\,5039			& LS\,I\,+61\degr303 	&  PSR\,B1259-63 	& HESS\,J0632+057 	& 1FGL\,J1018.6-5856	& LMC P3  \\
\midrule
\multicolumn{5}{l}{{\em HE surveys} (\%)}\\
FAVA  			& 0.9$\pm0.2$ 		&  4.0$\pm0.4$		& 10.7$\pm0.6$ 		& 0.1$\pm$0.1 		& 16.7$\pm0.7$		& 35.3$\pm1.0$ \\
3FGL~~~~~~~~~~~~~~~~& 70.2$\pm0.9$ 	& 39.9$\pm1.0$		& 15.7$\pm0.7$ 		& 0.3$\pm$0.1 		& 91.2$\pm0.6$		& 99.7$\pm0.1$ \\
\midrule
\multicolumn{5}{l}{{\em VHE surveys} (\%)}\\
HESS  			& 7.5$\pm0.5$ 		&  5.7$\pm0.5$		& 1.6$\pm0.3$ 		& 0.7$\pm0.2$ 		& 7.0$\pm0.5$			& $-$ \\
HAWC 			& 5.2$\pm0.4$ 		&  4.4$\pm0.4$		& 0.8$\pm0.2$ 		& 0.3$\pm0.1$ 		& 5.4$\pm0.4$			& $-$ \\
CTA  			& 50.4$\pm1.0$ 		&  38.3$\pm1.0$		& 10.8$\pm0.6$ 		& 3.7$\pm0.4$ 		& 47.8$\pm1.0$		& $-$ \\
\bottomrule
\end{tabular}
\end{table*}

\section{A synthetic population\label{sec:pop}}

In the preceding section we estimated  the number of existing gamma-ray binaries from the properties of the known systems. However, these systems represent only the upper end of the luminosity function of gamma-ray binaries. In this section we will estimate this number from a synthetic population model. Building this population requires a model for the gamma-ray emission of binaries, a bold enterprise given current knowledge. While there is general agreement that anisotropic inverse Compton scattering of photons from the star and $\gamma\gamma$ pair production at TeV energies must play a role, since these processes naturally lead to orbital modulations, the details vary significantly from model to model. Modulated Doppler boosting is also very likely to intervene if the emission occurs in a pulsar wind bow shock. Reproducing the orbital phases of gamma-ray detections in systems with Be companions such as \psrb\ has proven particularly difficult, possibly because of the complex interaction between the pulsar and the circumstellar material surrounding its companion. In the following, we adopted a simple model with the intention of minimizing the number of parameters while still being able to produce orbital lightcurves comparable to the observed ones.

\subsection{Orbital parameters\label{sec:ecc}}
The binary eccentricities $e$ were assumed to follow the thermal distribution \citep{ambartsumian} $p(e)de=2ede$ with the additional conditions that $e<e_{\rm max}=1-(P_{\rm orb}/ 2\rm\,days)^{-2/3}$ to ensure that the companion does not fill more than 70\% of its Roche lobe at periastron and that the binaries are circularised ($e=0$) for $P_{\rm orb} \leq 2$\,days (see \citealt{2017ApJS..230...15M} and references therein). The inclination of the system is derived by randomly picking a vector on a sphere. The argument of periastron and the orbital phase at the time of the first simulated observation are picked from a uniform distribution between 0 and $2\pi$. Finally, we uniformly sampled the logarithm of orbital periods between 1 and $10^{4}$ days in order to assess the fraction of  detected systems as a function of $P_{\rm orb}$, except in \S\ref{sec:full} where this is slightly modified for a more realistic representation of the $P_{\rm orb}$ distribution of HMXBs.

\subsection{Radiation model\label{sec:rad}}
We assume that the radiation is due to Compton upscattering of stellar photons with an initial energy $\approx 10\,\rm eV$. GeV (resp. TeV) emission  then requires electrons with a Lorentz factor $\gamma= 10^{4}$ (resp. $\gamma= 10^{6}$). For simplicity, we assumed monoenergetic distributions at these energies. This is supported by the observed GeV spectra of gamma-ray binaries, which generally consist of a hard power-law with an exponential cutoff around 1 GeV. This is also admissible in the TeV range where soft power laws are observed so that most of the photons have an energy close the threshold energy of the VHE observations. The true particle distributions are likely to be more complex but the GeV and TeV emissions are dominated by electrons of these energies and assuming more complex distributions (power laws, see \S5) does not have a significant impact on the results.

We computed the inverse Compton bolometric power radiated by these particles, assuming that they are located at the position of the compact object. If the electron distribution is isotropic, the lightcurve in the Thomson approximation for Compton scattering is given by 
\begin{equation}
L_\gamma=N_e \sigma_T c U_\star (1-\beta\mu) \left[(1-\beta\mu)\gamma^2-1\right]
\label{star}
\end{equation}
where $U_\star=(1/c)  \sigma_{SB} T^4_\star (R_\star/d_\star)^2$ with $T_\star$ the star temperature, $R_\star$ its radius, $d_\star$ its distance to the particles, and $\mu=\cos\theta$ represents the angle between the line-of-sight and the binary axis. The angle $\theta$ varies from $\pi/2+i$ (superior conjunction) to $\pi/2-i$ (inferior conjunction) with $i$ the system inclination. The massive star was assumed to have a radius of 10\,R$_{\odot}$ and temperature of 33,000\,K. The analytic formula is valid for $\gamma=10^4$, where the Thomson approximation is acceptable. However, stellar photons scatter in the Klein-Nishina regime when $\gamma=10^6$. Hence, we numerically integrated the Compton kernel to derive the anisotropic emitted power  instead of using Eq.~\ref{star} (see \citealt{2010A\string&A...516A..18D}). 

\begin{figure}
\begin{center}
\includegraphics[width=0.8\linewidth]{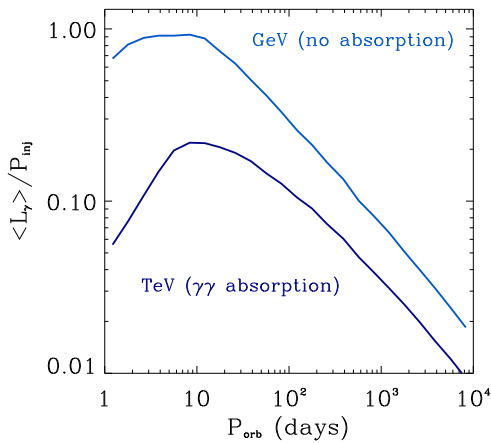} 
\caption{Mean orbit-averaged gamma-ray flux, normalized to the injected power, as a function of $P_{\rm orb}$ (see \S\ref{sec:rad}). }
\label{fig:power}
\end{center}
\end{figure}

The total number of electrons $N_{e}$ is related to the injected power in particles $P_{\rm inj}$ by
\begin{equation}
N_{e}=\frac{P_{\rm inj}}{\gamma m_{e} c^{2}} \times \mathrm{min}\left\{ \tau_{\rm esc},\tau_{\rm ic}\right\}
\end{equation}
where  $\tau_{\rm esc}$ is the escape timescale of the particles from the gamma-ray emitting region (see below) and $\tau_{\rm ic}$ is the inverse Compton loss timescale, which in the Thomson regime is 
\begin{equation}
 \tau_{\rm ic}=\frac{\gamma m_e c^2}{\tfrac{4}{3}\sigma_T c U_\star \gamma^2}.
\end{equation}
Hence, $<L_\gamma>=P_{\rm inj}$ (integrated over all angles) if the particles radiate efficiently before they leave the vicinity of the star ($\tau_{\rm ic}\leq \tau_{\rm esc}$), otherwise the radiated power is reduced to the fraction of particles that are in the emission zone $<L_\gamma>=(\tau_{\rm esc}/\tau_{\rm ic}) P_{\rm inj}$. Note that the latter can be rewritten using Kepler's third law as $L_\gamma \propto 1/d_{\star} \propto P_{\rm orb}^{-2/3}$, hence there is a break in the distribution of $<L_\gamma>/P_{\rm inj}$ as a function of $P_{\rm orb}$ for the orbital period where $\tau_{\rm ic}=\tau_{\rm esc}$. This can be seen in Figure~\ref{fig:power} where the mean of the average orbital  luminosity is plotted for a sample of $10^4$ binaries with orbital periods ranging from 1 to $10^4$ days and randomly sampled eccentricities. The break is at $P_{\rm orb}\approx 10$\,days because we decided to set $\tau_{\rm esc}=d_{\star}/c$. Such a fast escape timescale is reasonable in the context of gamma-ray binaries, where the accelerated particles flow away relativistically in a bow shock \citep{2015A\string&A...581A..27D}. This assumption is also conservative in that it may underestimate the number of detections by minimizing the radiative efficiency. The influence of this choice on the results is further discussed in \S\ref{sec:uncertainty}.

Since TeV photons are likely to create pairs by interacting with photons from the star, we calculated the expected $\gamma\gamma$ absorption at 1 TeV in the point source limit following \citet{2006A\string&A...451....9D}. Absorption reduces the average TeV luminosity for short orbital period systems, where the stellar radiation density is highest, resulting in a strong decrease of $L_{\rm TeV}$ with $P_{\rm orb}$ below 10 days (Fig.~\ref{fig:power}). GeV emission is not affected by $\gamma\gamma$ absorption. However, we also took into account eclipses of the (point-like) gamma-ray emission zone by the star. This results in a slight decrease of the average GeV power at short $P_{\rm orb}$, instead of the expected flat distribution $<L_\gamma>=P_{\rm inj}$. Hence,  this model predicts the radiative efficiency is  maximum for systems with $P_{\rm orb}\approx 10\rm\,days$.

\begin{figure} 
\begin{center}
\includegraphics[width=0.8\linewidth]{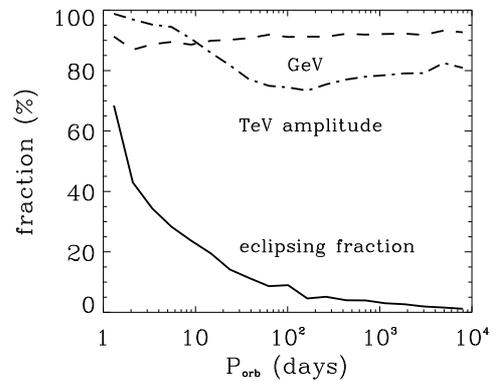} 
\caption{Mean fractional amplitude of the simulated GeV (dashed line) and TeV (dash-dotted line) gamma-ray lightcurves (see Fig.~\ref{fig:power}) and fraction of the systems showing eclipses as a function of $P_{\rm orb}$ (thick line).}
\label{fig:amplitude}
\end{center}
\end{figure}

Figure~\ref{fig:amplitude} shows the average fractional amplitude of the model TeV lightcurves, measured as $(f_{\rm max}-f_{\rm min})/(f_{\rm max}+f_{\rm min})$, where $f$ is the flux. The mean amplitude increases slightly  from short to long orbital periods due to the larger eccentricities permitted (see \S \ref{sec:ecc}) but eclipses and $\gamma\gamma$ absorption strongly increase the amplitude at short $P_{\rm orb}$. The average TeV variability amplitude at long  $P_{\rm orb}$ is about 80\%, implying $f_{\rm min}\approx 0.11 f_{\rm max}$. Fig.~\ref{fig:examples} shows examples of GeV and TeV lightcurves computed using the radiative model described in this section.

\subsection{Detectable fraction based on synthetic lightcurves\label{sec:synth}}
To check for consistency with the results of Table~\ref{table:detectable_grbi}, we produced $10^4$ synthetic lightcurves using the orbital period and eccentricity (when known) for each observed system {\em i.e.} leaving the system orientation free. We then normalized the synthetic lightcurves to the maximum observed luminosity. The systems are distributed throughout the Galaxy. The detection fractions in Table~\ref{table:synthetic_grbi} are within a factor 2 or less of those in Table~\ref{table:detectable_grbi}, showing comparable trends when looking at objects, orbital period or surveys. The exception is \psrb\ where the detected fraction in the 3FGL-like survey is a factor 5 higher because the model typically produces a lower amplitude lightcurve than observed, hence a higher average flux (see below). Despite this shortcoming, our simple radiative model should still be able to yield realistic estimates of the average detection rate for a population of systems.

\begin{figure}
\begin{center}
\includegraphics[width=0.495\linewidth]{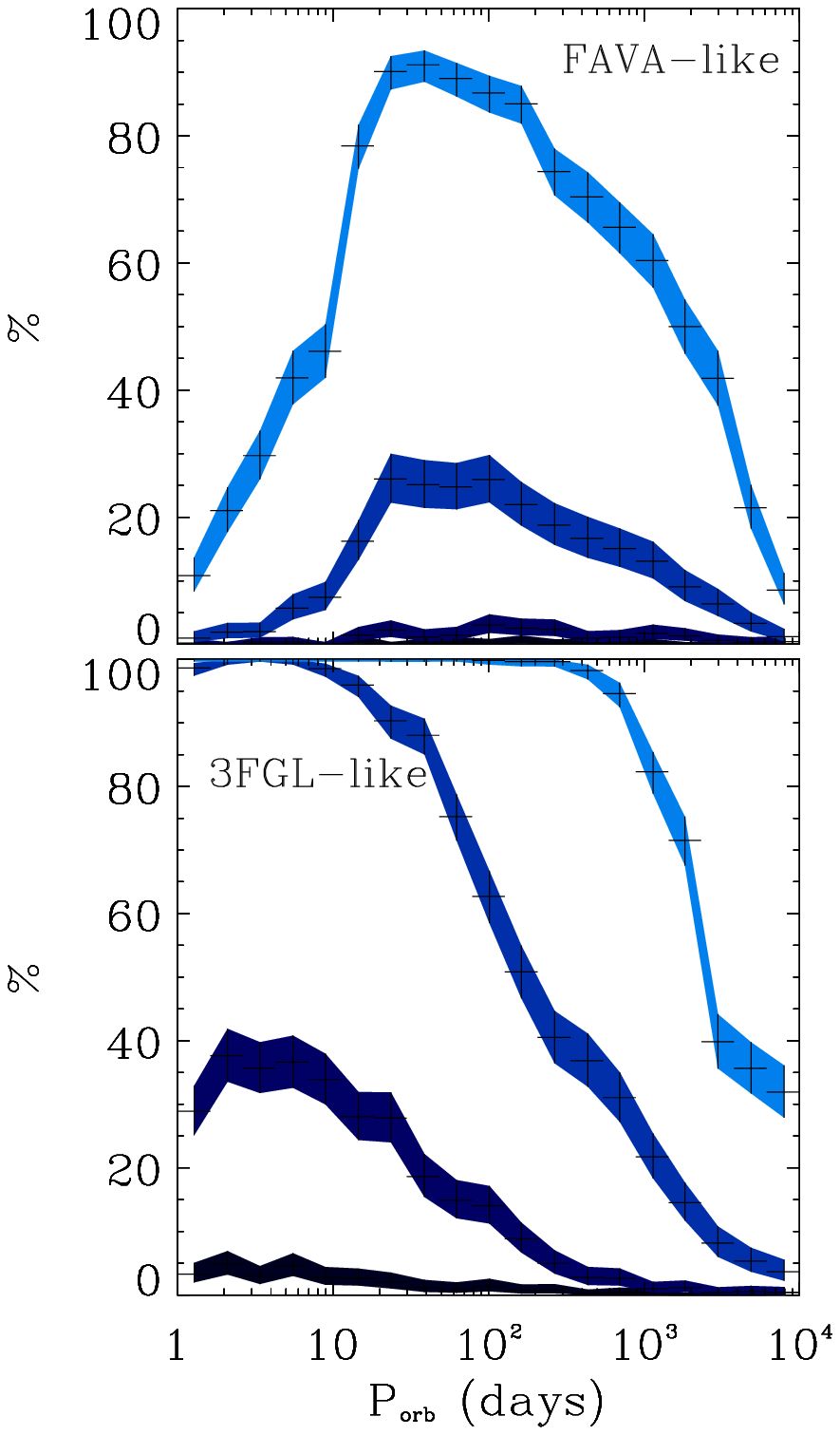}
\includegraphics[width=0.495\linewidth]{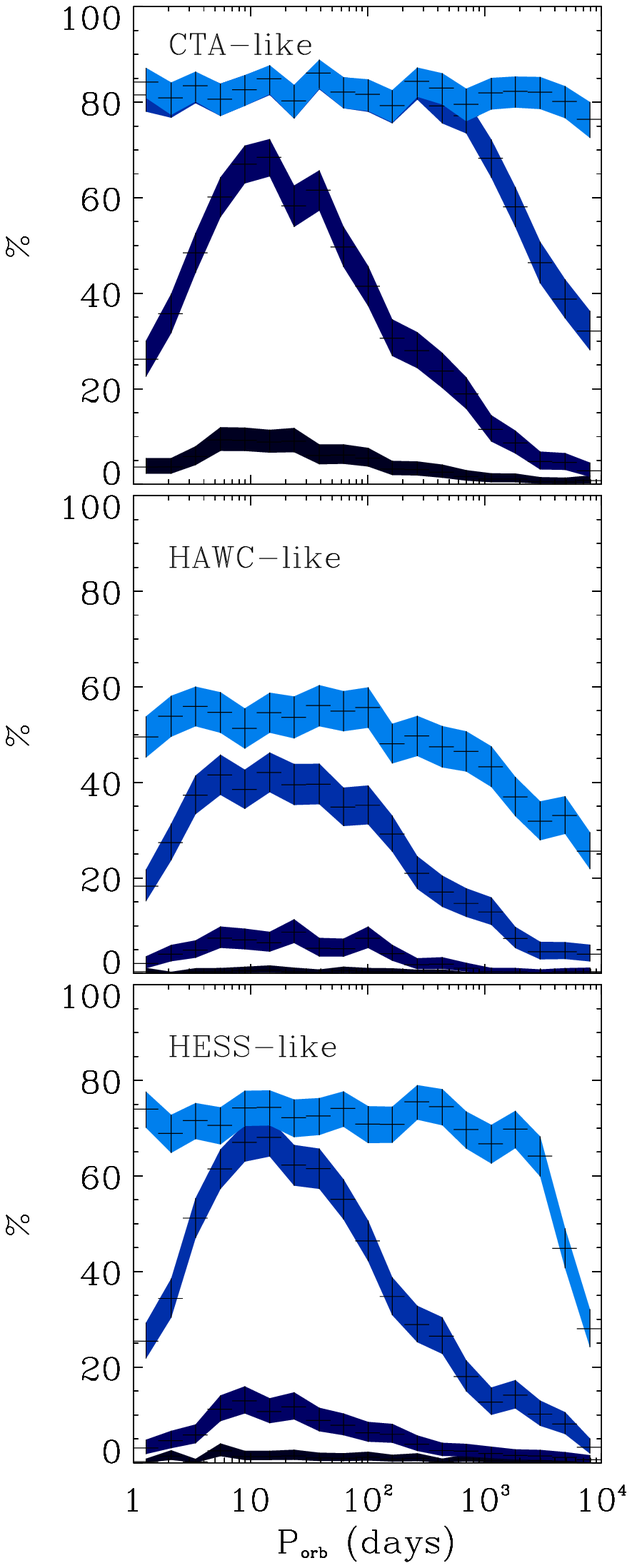}
\caption{Detected fractions in the HE (left panels) and VHE (right panels) surveys. Each panel contains four curves corresponding to $P_{\rm inj}=10^{33},10^{34},10^{35},10^{36}$ erg/s (dark to light blue sequence in each panel). The detection fraction increases when going from small $P_{\rm inj}$ (darker blue) to high $P_{\rm inj}$ (lighter blue) until the detected fraction saturates at the value given by the observable fraction (Tab.~\ref{tab:observable}).}
\label{fig:detect}
\end{center}
\end{figure}

We then produced synthetic lightcurves for a sample of binaries with random orbital parameters and a given injected power. Figure~\ref{fig:detect} shows the fraction of systems detected in the mock HE and VHE surveys discussed in \S\ref{sec:surveys}, as a function of $P_{\rm orb}$ and $P_{\rm inj}$. The FAVA-like survey is much less efficient at detecting systems than the 3FGL-like survey. At short orbital periods, the sensitivity is insufficient to detect systems on a timespan of a week. At long orbital periods, the amplitude of the variations in the model lightcurves (Fig.~\ref{fig:amplitude}) is insufficient to provide a significant advantage to this `burst' search strategy compared to the `integration' strategy employed in the 3FGL-like survey. The latter is extremely efficient when the injected power in HE-emitting particles exceeds $10^{35}$\,\ergs, even for long $P_{\rm orb}$ compared to the integration time (4 years). 

The VHE surveys access only part of the Galactic Plane so their maximum efficiency does not reach 100\% even for high injected powers in VHE-emitting particles. The results show comparable efficiencies for the HESS and HAWC-like surveys. The design of these two surveys, notably the visit frequencies, does not appear to play a major role in the detectable fraction: the peak at $P_{\rm orb}\approx 10$ to 100 days simply reflects the higher radiated luminosity expected for those orbital periods in the model (see Fig.~\ref{fig:power}). The CTA-like survey is much more sensitive, detecting nearly all accessible systems for $P_{\rm inj}\geq 10^{35}$\,\ergs\ regardless of orbital period. Again, the sensitivity at long  $P_{\rm orb}$ results from our model, which on average gives a minimum flux around 11\% of the maximum flux (Fig.~\ref{fig:amplitude} and \S\ref{sec:rad}). This enables the detection of long orbital period systems even when the phases of maximum flux are not sampled by the visits.

\subsection{Full population model\label{sec:full}}
A full population model requires assumptions on the injected power $P_{\rm inj}$ and how it relates to the total available power $\dot{E}$ measured by pulsar spindown. $P_{\rm inj}$ is likely to be different for the GeV and TeV emitting particles, whether they arise from different populations or from the same power-law distribution. We used \psrb, the only system with a measured $\dot{E}=8\times 10^{35}\rm\,erg\,s^{-1}$, to estimate the power going to the GeV and TeV-emitting particles. Simulating GeV and TeV lightcurves with the same orbital period and eccentricity as  \psrb\ (i.e. following the procedure described in \S\ref{sec:synth}), we found that injection fractions $P_{\rm GeV}=0.07 \dot{E}$ and $P_{\rm TeV}=0.01\dot{E}$ are needed to reproduce, on average, the peak gamma-ray fluxes listed in Tab.~\ref{table:detectable_grbi} and adopted these values in the following.

\begin{figure*}
\begin{center}
\includegraphics[width=0.7\linewidth]{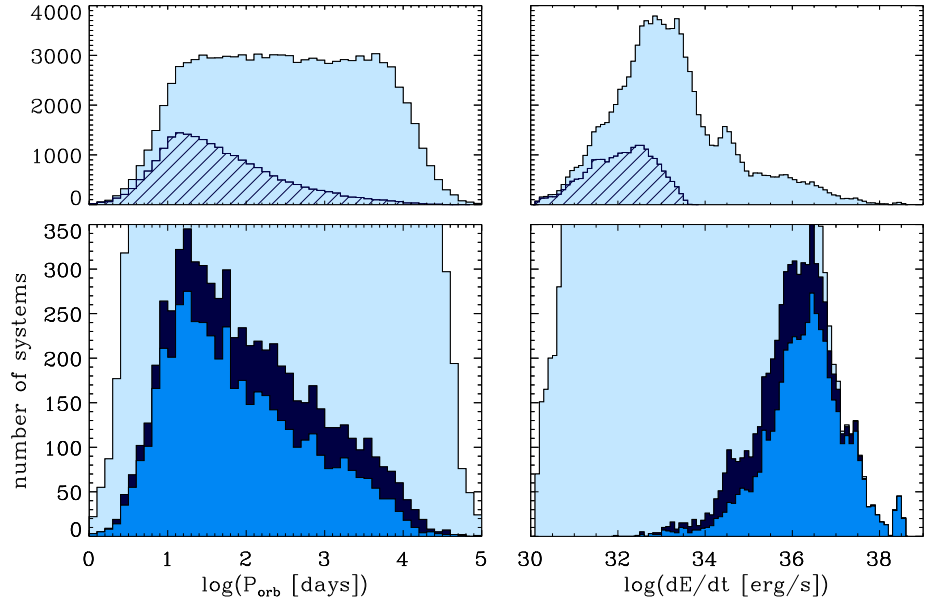}
\caption{Orbital period (left panels) and spindown power distribution (right panels) of a random sample of 10$^5$ systems. Top panels show the full population as a light blue histogram with the line-filled fraction showing the distribution of the binaries that are accreting according to Eq.~\ref{eq:acc}. The bottom panels zoom in to highlight the systems detected in any of the present-day surveys (combining the HESS, 3FGL and FAVA-like surveys, medium blue histogram) or any of the future surveys (combining the full CTA, extended 3FGL and HAWC-like surveys, dark blue histogram). The latter essentially shows the systems detected in any survey since a  binary detected in one of the present-day surveys has a nearly $>$99\% chance of being re-detected in one of the future surveys.}
\label{fig:synthetic}
\end{center}
\end{figure*}

The mock population is built by randomly sampling probability distributions of $\dot{E}$ and $P_{\rm orb}$. Following \citet{2013MNRAS.431..327L}, we took a flat distribution in $\log P_{\rm orb}$ tapered by gaussian edges at $\log P_{\rm orb}{\rm\,(days)}=1.3$ and 3.7. This probability distribution (Fig.~\ref{fig:synthetic}, top left panel) results from the evolution of pre-HMXB binaries \citep{2012ApJ...746...22B}. For $\dot{E}$, we took as input the distribution of spindown powers extracted from the ATNF pulsar catalogue \citep{2005AJ....129.1993M}\footnote{\href{http://www.atnf.csiro.au/people/pulsar/psrcat/}{http://www.atnf.csiro.au/people/pulsar/psrcat/}}, selecting only those pulsars with a pulse period $>$10\,ms and a spindown timescale $< 10^7$\,year to exclude recycled millisecond pulsars. The resulting $\dot{E}$ distribution is shown in the top right panel of Fig.~\ref{fig:synthetic}.

The detection fraction in the various surveys is calculated from a random sample of $10^5$ systems (Tab.~\ref{tab:detf}). The distributions of detected systems in present-day or future surveys are shown in the zoomed-in bottom panels of Fig.~\ref{fig:synthetic}.  The detection fractions are biased towards short $P_{\rm orb}$ and high $\dot{E}$, as expected from the results of \S\ref{sec:synth}. The population model naturally accounts for the existence of radio pulsars in binaries that remain undetected in gamma rays due to their long orbital periods and low spindown powers. PSR J0045-7319, PSR J1638-4725, PSR J1740-3052 with $P_{\rm orb}= 51,  1941, 231 $\,days (resp.) and $\dot{E}$=0.2, 0.4, and $5\times 10^{33}\rm\,erg\,s^{-1}$ (resp.) are examples of such systems that we do not expect to be readily detectable by the gamma-ray surveys \citep{2001MNRAS.325..979S,2011MNRAS.412L..63B,2012MNRAS.425.2378M}.

The pulsar wind pressure can be insufficient to hold off accretion from the companion's stellar wind for low $\dot{E}$ and short $P_{\rm orb}$, in which case we consider that the system will be an accreting HMXB (or a propeller, depending on the respective locations of the co-rotation radius and magnetospheric radius, both of which are within the light cylinder) and will not emit gamma rays. We used a simple criterion to test whether a system is accreting or not, assuming the massive star wind is isotropic, uniform with the same constant mass loss rate $\dot{M}_{\rm w}=10^{-6}\rm\,M_\odot\,yr^{-1}$ and  velocity $v_{\rm w}=1000\rm\,km\,s^{-1}$ for all systems. The system is accreting if the pulsar spindown power is less than 
\begin{equation}
\dot{E} < 4\cdot 10^{33} \left(\frac{\dot{M}_{\rm w}}{10^{-6}\rm\,M_\odot\,yr^{-1}}\right)\left(\frac{10^{3}\rm\,km\,s^{-1}}{v_{\rm w}}\right)^3\left(\frac{0.1\rm\,AU}{a_p}\right)^2\rm erg\,s^{-1}
\label{eq:acc}
\end{equation}
where $a_p$ is the binary separation at periastron \citep{1975A\string&A....39..185I}. This criterion is simplistic in regards to the complex physics of wind launching, capture, Be circumstellar discs etc \citep{2013A\string&ARv..21...64D}, but we have chosen values of $\dot{M}_{\rm w}$ and $v_{\rm w}$ that are likely to overestimate the fraction of accreting systems. We find about  23\% of the sampled systems are accreting, mostly at short $P_{\rm orb}$ and low $\dot{E}$ as shown by the line-filled histogram in the top panels of Fig.~\ref{fig:synthetic}. Despite this, we find negligible overlap with the population of systems detected in the HE and VHE surveys because these select high $\dot{E}$ systems ($<0.4\%$ of the detected systems are also flagged as accreting).

About 40\% of the binaries that are detected in HE can be found in both the 3FGL and FAVA-like surveys whereas less than 2\% are detected only in the FAVA-like survey. The detection of \psrb\ in FAVA without a concurrent detection in the 3FGL survey is therefore unlikely in our model, as the statistics in \S\ref{sec:synth} already showed. However, \psrb\ is close to our crude 3FGL detection threshold using the template lightcurve so details in the 3FGL  detectability may come into play  (orbit-to-orbit fluctuations, Galactic diffuse emission). The probability to detect a system in one of the VHE surveys (HESS, HAWC and CTA-like) is 4.23$\pm0.13$\%, with most of the detections arising from the CTA-like survey. Altogether, the probability to detect a gamma-ray binary in any of the surveys is 5.32$\pm0.14$\% (3FGL, FAVA, HESS, HAWC, or CTA-like). Only a very small number are detected in VHE surveys without a detection in the HE surveys with this model. For instance, all of the systems detected by the HESS-like survey are also detected by the 3FGL-like survey.

\begin{table}
\caption{Detection fractions for the population shown in Fig.~\ref{fig:synthetic}.}
\label{tab:detf}
\centering
\begin{tabular}{lc}
\toprule
\toprule
mock survey & detection fraction  (\%)\\
\midrule
3FGL or FAVA & 4.83$\pm0.13$\\
FAVA & 1.96$\pm0.09$\\
3FGL & 4.73$\pm0.13$\\
3FGL (extended) & 6.21$\pm0.15$\\
\midrule
HESS & 1.17$\pm0.07$ \\
HAWC & 0.88$\pm0.06$\\
CTA & 3.78$\pm0.12$\\
CTA (full) & 5.83$\pm0.15$\\
\bottomrule
\end{tabular}
\end{table}

\section{Discussion\label{sec:disc}}

\subsection{Estimated population of gamma-ray binaries}

We explored two ways to estimate the number of gamma-ray binaries. The first (\S\ref{sec:lc}) employed the lightcurves of the known systems as templates to evaluate the detection probability of identical systems distributed throughout the Galaxy. The results show that systems like \ls, \lsi, \dem\ and \fgl\ are already detectable throughout most of the Galaxy, so it is highly unlikely that more than 1 or 2 have escaped notice (Tab.~\ref{table:detectable_grbi}). Future HE detections are more likely to be of low duty cycle systems like \psrb, the total number of which is estimated at 7$^{+26}_{-6}$.  One such anticipated detection is that of PSR J2032+4127, a 2$\times10^{35}\rm\,erg\,s^{-1}$ pulsar in an eccentric, $>$20 year orbit around a Be star that will pass periastron in late 2017 \citep{2015MNRAS.451..581L,2017MNRAS.464.1211H}\footnote{PSR J2032+4127 is a {\em pulsed} {\em Fermi}-LAT source and coincident with an extended, persistent VHE source. It does not yet show evidence for variable gamma-ray emission related to binary motion, as seen in \psrb\ and the other gamma-ray binaries.}.  The largest source of uncertainty is the number of \hessj-like systems with a ratio of TeV  to GeV luminosity about two order-of-magnitude higher than the other binaries (see Tab.~\ref{table:detectable_grbi} and \citealt{2017arXiv170704280L}). There may be as many as $\approx$230 such systems in our Galaxy, an upper limit that CTA will decrease to $8^{+30}_{-7}$ if none are discovered in the full Galactic Plane survey (\S\ref{sec:lc}).

The known systems represent only the upper end of the luminosity function of gamma-ray binaries. Thus, our second estimate for the number of gamma-ray binaries employed a full population model based on a series of assumptions on the radiative process, the distributions of orbital parameters and injected power (\S\ref{sec:pop}). In the HE domain, with four systems in the 3FGL and FAVA surveys (Tab.~\ref{tab:detected}, excluding \dem\ since it is not in the Galactic Plane), the total parent population is estimated at 82$^{+108}_{-56}$ systems based on the detection fraction in Tab.~\ref{tab:detf}. In the VHE domain, with only \ls\ detected in the HESS survey, the parent population is constrained to 85$^{+290}_{-81}$ systems. Combining all the information in Tab.~\ref{tab:detected} into the likelihood function, {\em i.e.} assuming four systems in the HE surveys, one system in HESS, none in HAWC, and at least five in the full CTA survey, the population is estimated at 101$_{-52}^{+89}$ gamma-ray binaries in our Galaxy. These numbers are consistent with the predictions  from population synthesis of HMXBs (\S\ref{sec:intro}).

Gamma-ray surveys are $\geq50\%$ complete for $\dot{E}\geq 10^{36}\rm\,erg\,s^{-1}$ (Fig.~\ref{fig:synthetic}), but they access only a handful of systems in a population of about a hundred binaries. A few additional systems  like PSR J2032+4127 may be detected through their pulsed gamma-ray emission without showing binary-related gamma-ray emission. We have not attempted to take this into account.  The spindown distribution of detected {\em Fermi}-LAT pulsars peaks at $\log \dot{E}=35.5$ (see \S\ref{sec:uncertainty} below) suggesting this is unlikely to make a difference to the number of systems detected in gamma rays. A couple dozen binaries may be visible as accreting X-ray sources, indistinguishable from other HMXBs except perhaps through their neutron star spin periods or through propeller-induced behaviour. \object{SAX J0635+0533} \citep{2000ApJ...528L..25C} and \object{A0538-66} \citep{1982Natur.297..568S} are possible examples. These two systems clearly have much faster spin periods ($<70\,$ms) than all the other known X-ray pulsars in HMXBs ($>$ 1 to 1000s), suggesting that the neutron star may not  yet have spun down significantly from its birth period.  

PSR J0045-7319, PSR J1638-4725, and PSR J1740-3052 are representative of the low $\dot{E}$ systems that represent the  majority of the pulsar + massive star population: $\approx 55\%$ of the sampled systems have $\dot{E}\leq 10^{34}\rm\,erg\,s^{-1}$ and are not accreting. Adding in the 23\% that are accreting, this implies that 78\% of the population is inaccessible to gamma-ray surveys. Estimating their detection rate in radio (SKA) or X-ray surveys (eROSITA) is beyond the scope of this work but we note that the long $P_{\rm orb}$, high eccentricity systems are clearly more susceptible to be detected as radio pulsars \citep{1994A\string&A...282...61L}, providing a complementary way to access the pulsar + massive star population. 

\subsection{Systematic uncertainties in the population synthesis\label{sec:uncertainty}}

How dependent are our results on the assumptions of the model? The Galactic distribution and binary parameters should not be a major source of concern since these have already been scrutinized in population studies of high-mass X-ray binaries \citep{2015A\string&ARv..23....2W}. The distribution of $\dot{E}$ for gamma-ray binaries is entirely unknown and taking as input the $\dot{E}$ distribution of young pulsars in the ATNF catalog probably suffers from a variety of selection biases, notably because it is not obvious that the birth spin period and evolution should be identical in isolated pulsars and binaries (if only because  mass loss and kick during the supernova are necessarily weaker if the newly-born neutron star is to remain bound to its companion). Yet, our assumption on $\dot{E}$ is not likely to have a major impact on gamma-ray observations since these are mostly sensitive to the high end of this distribution. The $\dot{E}$ distribution of detected binaries (Fig.~\ref{fig:detect}) actually resembles the $\dot{E}$ distribution of young pulsars detected in gamma-rays with the {\em Fermi}-LAT\footnote{\href{https://confluence.slac.stanford.edu/display/GLAMCOG/Public+List+of+LAT-Detected+Gamma-Ray+Pulsars}{link} to the public list of LAT-detected gamma-ray pulsars}, which we find to be well-approximated by a Gaussian centered at $\log \dot{E}=35.5$ with a standard deviation $\sigma=1$. Taking this distribution as input increases the detection fraction, decreasing the population size inferred from current observations without changing much the number of expected detections in future surveys. However, this distribution cannot account for the known radio pulsars in orbit around massive stars with low $\dot{E}$. The strongest impact of our assumption on the $\dot{E}$ distribution is therefore  on the relative numbers of pulsar+massive star binaries that are found in radio and gamma-ray surveys. 

The lightcurves of gamma-ray binaries have proven difficult to model, even in the cases where we have the most information, questioning the validity of our radiative model. For instance, relativistic beaming of the  emission is thought to be an important factor in shaping the lightcurves (e.g. \citealt{2017ApJ...838..145A}). A  refined lightcurve model is desirable but may not change our results much. First, despite its simplicity, the detection fractions inferred from the model are broadly consistent with those inferred from the observed lightcurves (\S\ref{sec:template}). Its main shortcoming is that it predicts lower amplitudes than observed, overestimating detection rates for \psrb-like systems. However, these do not dominate the detected systems (Fig.~\ref{fig:synthetic}). Second, we have experimented with a more complex radiative model, using a power-law distribution of particles and including Doppler boosting (assuming a particle bulk velocity of c/3 directed away from the star as in \citealt{2010A\string&A...516A..18D}). There was surprisingly little difference with the detection fractions shown in Fig.~\ref{fig:detect} despite substantial changes to the lightcurves from relativistic Doppler boosting, indicating that the flux level is more important than the detailed shape of the lightcurve in setting the detection fractions. 

Improvements to our radiative model should thus concentrate on the injected power $P_{\rm inj}$  and the radiative efficiency $\tau_{\rm esc}/\tau_{\rm ic}$, both of which set the flux level. A longer escape timescale increases the radiative efficiency but this needs to be compensated by a lower fraction $P_{\rm inj}/\dot{E}$ in order to match the maximum flux from \psrb\ (\S\ref{sec:full}). For example, taking $\tau_{\rm esc}=10 d/c$ implies a decreased injection fraction ($P_{\rm GeV}\approx 0.02\dot{E}$ and $P_{\rm TeV}\approx 0.002\dot{E}$). The combination yields an estimated population of 105$_{-54}^{+92}$  systems, very close to our previous estimate of 101$_{-52}^{+89}$. The peak of the gamma-ray flux distribution is pushed to longer periods than in Fig.~\ref{fig:power}, leading to a flatter distribution in the fraction of detected systems as a function of $P_{\rm orb}$. In principle, the $P_{\rm orb}$ distribution of detected gamma-ray binaries could thus be used to constrain $\tau_{\rm esc}$, assuming excellent  knowledge of their parent $P_{\rm orb}$ distribution.  A distribution of $L_{\gamma}/\dot{E}$ as a function of $P_{\rm orb}$ would narrow down possibilities for the radiative and injection efficiency, presently only constrained by observations of \psrb. Gamma-ray observations of  PSR J2032+4127 at periastron passage will provide a second constrain on these efficiencies. Inversely, future observations of gamma-ray binaries as a population also have the potential to constrain the relative efficiencies in the GeV and TeV range, as described below (\S\ref{sec:future}).

Many of our results remain applicable even if gamma-ray binaries are not powered by pulsar spindown. The results of \S\ref{sec:extrapolating}, based on the template lightcurves, do not depend on this assumption. The results of \S\ref{sec:ecc}-\ref{sec:synth} are also applicable as long as the emission arises from electrons located close to the compact object (e.g. at the base of a jet) upscattering stellar radiation. Even if the compact object is a black hole, the compact object mass remains much lower than the companion mass so any  difference in orbits will be minor for the radiation model. However, differences can be expected in the full population model (\S\ref{sec:full}) since we have made use of the distribution of spindown powers of pulsars. We would need some assumption on the distribution of jet power to perform an equivalent calculation and deduce the parent population. However, as stated in the introduction, we consider it very unlikely that gamma-ray binary emission arises from accretion-powered jets \citep{2013A\string&ARv..21...64D}.

\subsection{Future gamma-ray observations\label{sec:future}}

What do future observations hold in store?  With an estimated population size of 101 gamma-ray binaries, up to 8 new binaries might be detected in an extended 3FGL survey with a most likely value of 2 new detections beyond the known sample. New discoveries are less likely in the VHE surveys. Once the expected detections (Tab.~\ref{tab:detected}) are taken into account, up to 3 new detections are predicted in the HAWC survey, 5 in the initial CTA survey and 6 in the full CTA survey (95\% confidence limits), the statistically most likely outcome being no new detection. The reason is that the detection probabilities remain small for these surveys. 

Serendipitous discoveries in deep VHE observations of Galactic sources (e.g. \hessj) can complement the surveys. We find that the  probability for a chance detection of a gamma-ray binary is 0.17$\pm0.03$\% in a 100 hour CTA exposure towards the Galactic Center, covering 6\degr\ in Galactic longitude, and reaching 1 mCrab at 1 TeV.  This is $\approx$1.7 times the detection rate from the Galactic Plane survey over a comparable area {\em i.e.} there are roughly 7 previously undetected systems for every 10 systems detected in the Galactic Plane survey of this deep field. Having 20 such deep pointings, spread around the Galactic Plane towards areas of special interest such as the Galactic Center, the Cygnus and Westerlund regions, or the Sagittarius-Carina spiral arm (see Fig.~\ref{fig:gal_distribution}), adds 1.4\% to the detected fraction with CTA. Combining surveys and deep pointings can thus yield  a  detection rate comparable to or greater than that in the {\em Fermi}-LAT survey.

Any discovery in a VHE survey would have a major impact on the estimated population number, raising it to higher values. A discrepancy could appear between the actual number of sources detected in the VHE and HE surveys since the model predicts that essentially all TeV sources should be detected at GeV energies.  Some tension is already present in the model. The maximum likelihood $L_m$ obtained by treating independently the HE and VHE surveys is $\geq 20\%$ for both, with corresponding population numbers of 82$^{+108}_{-56}$  (HE) and 132$^{+268}_{-86}$ (VHE). 
Combining the HE and VHE numbers into a single likelihood gives the estimate of 101$_{-52}^{+89}$ systems presented above, but $L_m$ drops to 4\%: this low probability indicates that the model has difficulty accounting for both the number of GeV and TeV detections when they are taken from the same underlying binary population. 
This can be resolved by increasing the injection fraction $P_{\rm inj}$ at 1 TeV, with the effect of raising the detection probability in VHE surveys and lowering the parent population size to a value that slackens the tension with the HE constrains, or by lowering it at 1 GeV with opposite effects on detection probability and population size. Hence, the relative numbers of HE and VHE detections can constrain the relative injection efficiencies.  
In any case, regardless of the value of $P_{\rm inj}$, the population of VHE-emitting systems is unlikely to be greater than 230 systems, otherwise \hessj-like systems would be detected in the 3FGL survey or in the HESS Galactic Plane survey (\S 3.3). This number is close to the upper limit on the population size estimated from synthetic lightcurves (190 systems). Both estimates thus converge to a maximum gamma-ray binary population of $\approx 200$ systems. With 200 systems, up to 10 (resp. 14) new binaries could be detected in the initial (resp. full) CTA survey, the most likely number being 4 (resp. 6) discoveries.

\section{Conclusions}

We have modelled the population of gamma-ray binaries and evaluated the fraction of systems that can be detected in various HE and VHE surveys, taking into account the variability of their gamma-ray emission. The number of gamma-ray binaries is constrained to 101$_{-52}^{+89}$ systems in our Galaxy. This number matches expectations from HMXB population synthesis. 

Gamma-ray binaries are rare systems and we do not expect a watershed of discoveries in the near future. Pursuing the {\em Fermi}-LAT survey to $\approx 2024$ should lead to a handful of discoveries, mostly \psrb-like systems. At very high energies, combining Galactic Plane surveys and deep observations of Galactic sources with CTA should provide a comparable number of discoveries. However, the number of \hessj-like systems, with very weak GeV emission, is a major source of uncertainty. Observations already indicate that the GeV and TeV emission  originate from different particle populations. A VHE survey could therefore reveal a population of binaries that cannot be seen with the {\em Fermi}-LAT. Such a population is limited to $\la$ 230 systems based on the lack of \hessj-like systems in the {\em Fermi}-LAT 3FGL survey and the HESS Galactic Plane Survey. With 200 systems, four new gamma-ray binaries  can be expected in the first two years of the CTA Galactic Plane survey. Of course, these numbers refer only to gamma-ray binaries and do not limit gamma-ray detections from other types of binaries such as novae, colliding wind binaries, binary millisecond pulsars, microquasars, etc.

Detecting a system depends more on its orbit-averaged flux than on the shape of the gamma-ray lightcurve. Thus, the scheduling of visits from ground-based instruments plays a minor role in setting the detected fraction. The average flux is set by the efficiency with which spindown power is radiated in the HE and VHE bands. This is the most important source of uncertainty in our model. Ideally, this should be constrained by measuring the pulsar spindown power and radiated luminosity for as many systems as possible. At present, this is limited to \psrb, with the possible addition of PSR J2032+4127 in the near future. Alternatively,  this relative efficiency in the HE and VHE bands can be constrained statistically by the relative number of sources detected in HE and VHE surveys. 

About 55\% of pulsars in orbit around massive stars are hardly accessible to gamma-ray observations, which are most sensitive to the high $\dot{E}$, short $P_{\rm orb}$ systems. Low $\dot{E}$ and long $P_{\rm orb}$ binaries are likely to be more efficiently accessed by radio pulsar surveys, which are thus fully complementary to the gamma-ray observations.  Another significant fraction, $\approx 23\%$, may actually be visible as accreting X-ray pulsars or propellers instead of binary pulsar wind nebulae.  Future work should strive to combine detection probabilities in gamma rays with detection probabilities in radio (SKA) and X-ray (eROSITA) surveys.

\begin{acknowledgement}
We are grateful to Masha Chernyakova, Markus B\"ottcher, and Jamie Holder for their careful CTA internal review that helped improve this work. GD and POP acknowledge support from {\em Centre National d'Etudes Spatiales} (CNES). 
\end{acknowledgement}
This paper has gone through internal review by the CTA Consortium.

\begin{appendix}
\section{Example synthetic lightcurves}
\begin{figure*}
\begin{center}
\includegraphics[width=\linewidth]{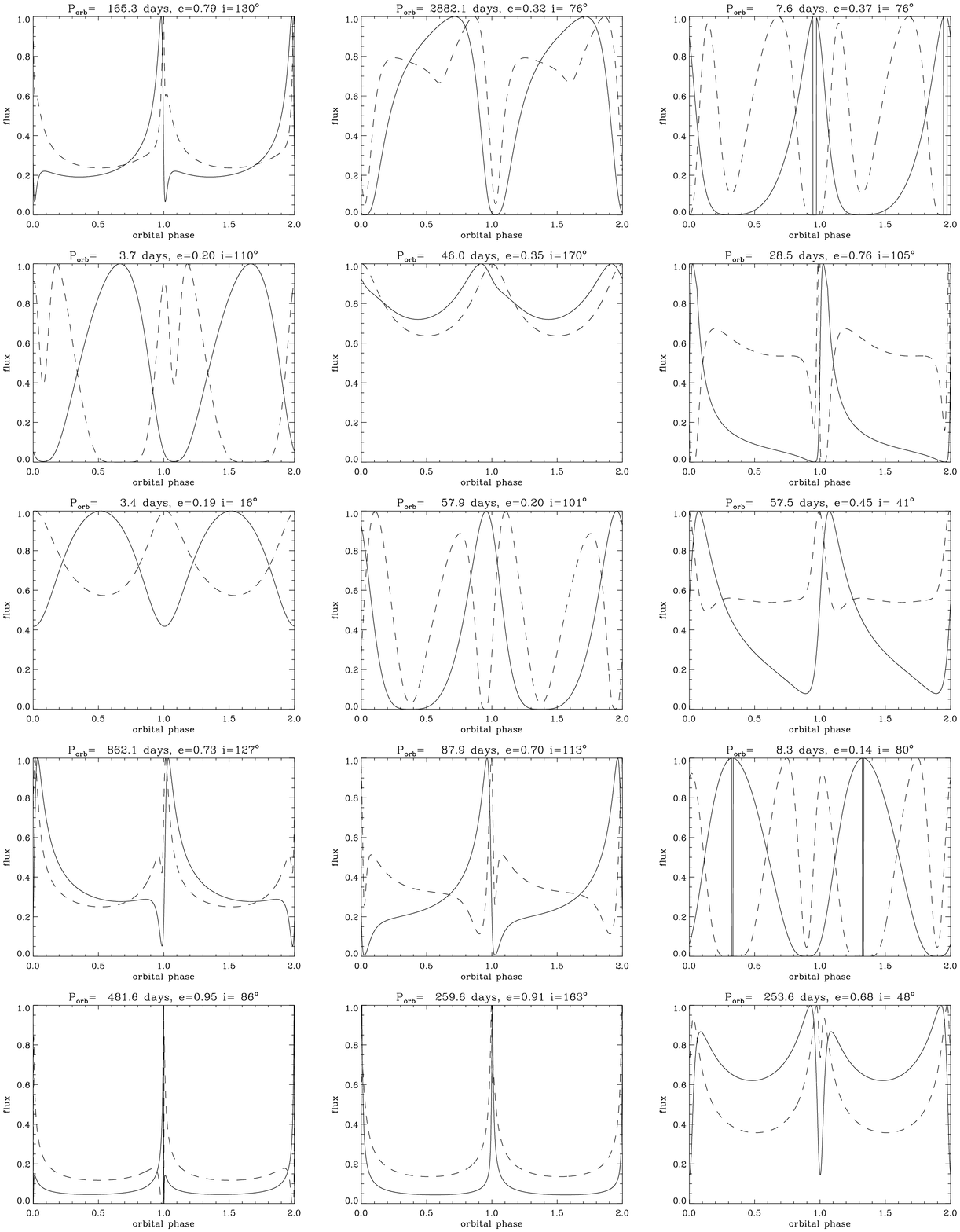}
\caption{Example lightcurves computed from the model described in \S\ref{sec:rad} (full line: GeV emission; dashed line: TeV emission, taking into account $\gamma\gamma$ absorption). The orbital period and eccentricity of the binary system is indicated in the title of each plot. The lightcurves are normalised to the maximum value. The systems shown here are a random selection of the systems flagged as detected in Fig.~\ref{fig:synthetic}. Periastron passage is at phase 0.}
\label{fig:examples}
\end{center}
\end{figure*}

\end{appendix}
\bibliographystyle{aa}
\bibliography{population_arxiv}

\begin{thebibliography}{51}
\expandafter\ifx\csname natexlab\endcsname\relax\def\natexlab#1{#1}\fi

\bibitem[{{Abdo} {et~al.}(2009)}]{2009ApJ...706L..56A}
{Abdo,  A.~A., {et~al.}  ({\em Fermi}-LAT collaboration)} 2009, \apjl, 706, L56

\bibitem[{{Abdollahi} {et~al.}(2016)}]{2016arXiv161203165A}
{{Abdollahi}, S., {et~al.} ({\em Fermi}-LAT collaboration)} 2016, ApJS, submitted
  [\eprint[arXiv]{1612.03165}]

\bibitem[{{Abeysekara} {et~al.}(2017)}]{2017ApJ...843...40A}
{Abeysekara, A.~U., {et~al.} (HAWC collaboration)} 2017, \apj, 843, 40

\bibitem[{{Abramowski} {et~al.}(2015)}]{2015A&A...577A.131H}
{Abramowski, A., {et~al.} (HESS collaboration)} 2015, \aap, 577, A131

\bibitem[{{Acciari} {et~al.}(2011)}]{2011ApJ...738....3A}
{Acciari, V.~A., {et~al.} (VERITAS collaboration)} 2011, \apj, 738, 3

\bibitem[{{Acero} {et~al.}(2015)}]{2015ApJS..218...23A}
{Acero, F., {et~al.} ({\em Fermi}-LAT collaboration)}. 2015, \apjs, 218, 23

\bibitem[{{Ackermann} {et~al.}(2013)}]{2013ApJ...773L..35A}
{Ackermann, M., {et~al.}  ({\em Fermi}-LAT collaboration)}. 2013, \apjl, 773, L35

\bibitem[{{Ackermann} {et~al.}(2012)}]{2012Sci...335..189F}
{Ackermann, M. {et~al.} ({\em Fermi}-LAT collaboration)} 2012, Science, 335, 189

\bibitem[{{Aharonian} {et~al.}(2006)}]{2006A&A...460..743A}
{Aharonian, F.~A., {et~al.} (HESS collaboration)} 2006, \aap, 460, 743

\bibitem[{{Albert} {et~al.}(2008)}]{2008ApJ...674.1037A}
{Albert, J., {et~al.} (MAGIC collaboration)} 2008, \apj, 674, 1037

\bibitem[{{Aliu} {et~al.}(2014)}]{2014ApJ...780..168A}
{Aliu, E., {et~al.} (VERITAS and HESS collaborations)} 2014, \apj,
  780, 168

\bibitem[{Ambartsumian(1937)}]{ambartsumian}
Ambartsumian, V.~A. 1937, Astron. Zhurn., 14, 207

\bibitem[{{An} \& {Romani}(2017)}]{2017ApJ...838..145A}
{An}, H. \& {Romani}, R.~W. 2017, \apj, 838, 145

\bibitem[{{Bassa} {et~al.}(2011){Bassa}, {Brisken}, {Nelemans}, {Stairs},
  {Stappers}, \& {Kramer}}]{2011MNRAS.412L..63B}
{Bassa}, C.~G., {Brisken}, W.~F., {Nelemans}, G., {et~al.} 2011, \mnras, 412,
  L63

\bibitem[{{Bhadkamkar} \& {Ghosh}(2012)}]{2012ApJ...746...22B}
{Bhadkamkar}, H. \& {Ghosh}, P. 2012, \apj, 746, 22

\bibitem[{{Bordas} {et~al.}(2016){Bordas}, {Dubus}, {Eger}, {Ernenwein},
  {Laffon}, {Mariaud}, {Murach}, {de Naurois}, {Romoli}, {Sch{\"u}ssler},
  {Zanin}, \& {for the H.~E.~S.~S.~Collaboration}}]{2016arXiv161003264B}
{Bordas}, P., {Dubus}, G., {Eger}, P., {et~al.}, 2017, in {6th International 
Meeting on High-energy gamma-ray astronomy}, 
{AIP Conference Proceedings, 1792, 040017} 

\bibitem[{{Campana} {et~al.}(1995){Campana}, {Stella}, {Mereghetti}, \&
  {Colpi}}]{1995A&A...297..385C}
{Campana}, S., {Stella}, L., {Mereghetti}, S., \& {Colpi}, M. 1995, \aap, 297,
  385

\bibitem[{{Carrami{\~n}ana}(2016)}]{2016JPhCS.761a2034C}
{Carrami{\~n}ana}, A. 2016, Journal of Physics Conference Series, 761, 012034

\bibitem[{{Corbet} {et~al.}(2016){Corbet}, {Chomiuk}, {Coe}, {Coley}, {Dubus},
  {Edwards}, {Martin}, {McBride}, {Stevens}, {Strader}, {Townsend}, \&
  {Udalski}}]{2016ApJ...829..105C}
{Corbet}, R.~H.~D., {Chomiuk}, L., {Coe}, M.~J., {et~al.} 2016, \apj, 829, 105

\bibitem[{{Cusumano} {et~al.}(2000){Cusumano}, {Maccarone}, {Nicastro},
  {Sacco}, \& {Kaaret}}]{2000ApJ...528L..25C}
{Cusumano}, G., {Maccarone}, M.~C., {Nicastro}, L., {Sacco}, B., \& {Kaaret},
  P. 2000, \apjl, 528, L25

\bibitem[{{Dubus}(2006)}]{2006A&A...451....9D}
{Dubus}, G. 2006, \aap, 451, 9

\bibitem[{{Dubus}(2013)}]{2013A&ARv..21...64D}
{Dubus}, G. 2013, \aapr, 21, 64

\bibitem[{{Dubus} {et~al.}(2010){Dubus}, {Cerutti}, \&
  {Henri}}]{2010A&A...516A..18D}
{Dubus}, G., {Cerutti}, B., \& {Henri}, G. 2010, \aap, 516, A18

\bibitem[{{Dubus} {et~al.}(2015){Dubus}, {Lamberts}, \&
  {Fromang}}]{2015A&A...581A..27D}
{Dubus}, G., {Lamberts}, A., \& {Fromang}, S. 2015, \aap, 581, A27

\bibitem[{{Gregory} \& {Taylor}(1978)}]{1978Natur.272..704G}
{Gregory}, P.~C. \& {Taylor}, A.~R. 1978, \nat, 272, 704

\bibitem[{{Grimm} {et~al.}(2002){Grimm}, {Gilfanov}, \&
  {Sunyaev}}]{2002A&A...391..923G}
{Grimm}, H.-J., {Gilfanov}, M., \& {Sunyaev}, R. 2002, \aap, 391, 923

\bibitem[{{Hadasch} {et~al.}(2012){Hadasch}, {Torres}, {Tanaka}, {Corbet},
  {Hill}, {Dubois}, {Dubus}, {Glanzman}, {Corbel}, {Li}, {Chen}, {Zhang},
  {Caliandro}, {Kerr}, {Richards}, {Max-Moerbeck}, {Readhead}, \&
  {Pooley}}]{2012ApJ...749...54H}
{Hadasch}, D., {Torres}, D.~F., {Tanaka}, T., {et~al.} 2012, \apj, 749, 54

\bibitem[{{Hinton} {et~al.}(2009){Hinton}, {Skilton}, {Funk}, {Brucker},
  {Aharonian}, {Dubus}, {Fiasson}, {Gallant}, {Hofmann}, {Marcowith}, \&
  {Reimer}}]{2009ApJ...690L.101H}
{Hinton}, J.~A., {Skilton}, J.~L., {Funk}, S., {et~al.} 2009, \apjl, 690, L101

\bibitem[{{Ho} {et~al.}(2017){Ho}, {Ng}, {Lyne}, {Stappers}, {Coe}, {Halpern},
  {Johnson}, \& {Steele}}]{2017MNRAS.464.1211H}
{Ho}, W.~C.~G., {Ng}, C.-Y., {Lyne}, A.~G., {et~al.} 2017, \mnras, 464, 1211

\bibitem[{{Iben} {et~al.}(1995){Iben}, {Tutukov}, \&
  {Yungelson}}]{1995ApJS..100..217I}
{Iben}, I.~J., {Tutukov}, A.~V., \& {Yungelson}, L.~R. 1995, \apjs, 100, 217

\bibitem[{{Illarionov} \& {Sunyaev}(1975)}]{1975A&A....39..185I}
{Illarionov}, A.~F. \& {Sunyaev}, R.~A. 1975, \aap, 39, 185

\bibitem[{{Johnston} {et~al.}(1992){Johnston}, {Manchester}, {Lyne}, {Bailes},
  {Kaspi}, {Qiao}, \& {D'Amico}}]{1992ApJ...387L..37J}
{Johnston}, S., {Manchester}, R.~N., {Lyne}, A.~G., {et~al.} 1992, \apjl, 387,
  L37

\bibitem[{{Li} {et~al.}(2017){Li}, {Torres}, {Cheng}, {de Ona Wilhelmi},
  {Kretschmar}, {Hou}, \& {Takata}}]{2017arXiv170704280L}
{Li}, J., {Torres}, D.~F., {Cheng}, K.-S., {et~al.} 2017, \apj, accepted  
  [\eprint[arXiv]{1707.04280}]

\bibitem[{{Lipunov} {et~al.}(1994){Lipunov}, {Nazin}, {Osminkin}, \&
  {Prokhorov}}]{1994A&A...282...61L}
{Lipunov}, V.~M., {Nazin}, S.~N., {Osminkin}, E.~Y., \& {Prokhorov}, M.~E.
  1994, \aap, 282, 61

\bibitem[{{Lutovinov} {et~al.}(2013){Lutovinov}, {Revnivtsev}, {Tsygankov}, \&
  {Krivonos}}]{2013MNRAS.431..327L}
{Lutovinov}, A.~A., {Revnivtsev}, M.~G., {Tsygankov}, S.~S., \& {Krivonos},
  R.~A. 2013, \mnras, 431, 327

\bibitem[{{Lyne} {et~al.}(2015){Lyne}, {Stappers}, {Keith}, {Ray}, {Kerr},
  {Camilo}, \& {Johnson}}]{2015MNRAS.451..581L}
{Lyne}, A.~G., {Stappers}, B.~W., {Keith}, M.~J., {et~al.} 2015, \mnras, 451,
  581

\bibitem[{{Madsen} {et~al.}(2012){Madsen}, {Stairs}, {Kramer}, {Camilo},
  {Hobbs}, {Janssen}, {Lyne}, {Manchester}, {Possenti}, \&
  {Stappers}}]{2012MNRAS.425.2378M}
{Madsen}, E.~C., {Stairs}, I.~H., {Kramer}, M., {et~al.} 2012, \mnras, 425,
  2378

\bibitem[{{Malyshev} \& {Chernyakova}(2016)}]{2016MNRAS.463.3074M}
{Malyshev}, D. \& {Chernyakova}, M. 2016, \mnras, 463, 3074

\bibitem[{{Manchester} {et~al.}(2005){Manchester}, {Hobbs}, {Teoh}, \&
  {Hobbs}}]{2005AJ....129.1993M}
{Manchester}, R.~N., {Hobbs}, G.~B., {Teoh}, A., \& {Hobbs}, M. 2005, \aj, 129,
  1993

\bibitem[{{Massi} {et~al.}(2017){Massi}, {Migliari}, \&
  {Chernyakova}}]{2017arXiv170401335M}
{Massi}, M., {Migliari}, S., \& {Chernyakova}, M. 2017, MNRAS, in press
  [\eprint[arXiv]{1704.01335}]

\bibitem[{{Meurs} \& {van den Heuvel}(1989)}]{1989A&A...226...88M}
{Meurs}, E.~J.~A. \& {van den Heuvel}, E.~P.~J. 1989, \aap, 226, 88

\bibitem[{{Moe} \& {Di Stefano}(2017)}]{2017ApJS..230...15M}
{Moe}, M. \& {Di Stefano}, R. 2017, \apjs, 230, 15

\bibitem[{{Paredes} {et~al.}(2000){Paredes}, {Mart{\'{\i}}}, {Rib{\'o}}, \&
  {Massi}}]{2000Sci...288.2340P}
{Paredes}, J.~M., {Mart{\'{\i}}}, J., {Rib{\'o}}, M., \& {Massi}, M. 2000,
  Science, 288, 2340

\bibitem[{{Portegies Zwart} \& {Verbunt}(1996)}]{1996A&A...309..179P}
{Portegies Zwart}, S.~F. \& {Verbunt}, F. 1996, \aap, 309, 179

\bibitem[{{Portegies Zwart} \& {Yungelson}(1998)}]{1998A&A...332..173P}
{Portegies Zwart}, S.~F. \& {Yungelson}, L.~R. 1998, \aap, 332, 173

\bibitem[{{Ringermacher} \& {Mead}(2009)}]{2009MNRAS.397..164R}
{Ringermacher}, H.~I. \& {Mead}, L.~R. 2009, \mnras, 397, 164

\bibitem[{{Russeil}(2003)}]{2003A&A...397..133R}
{Russeil}, D. 2003, \aap, 397, 133

\bibitem[{{Shvartsman}(1971)}]{1971SvA....15..342S}
{Shvartsman}, V.~F. 1971, Soviet Astronomy, 15, 342

\bibitem[{{Skinner} {et~al.}(1982){Skinner}, {Bedford}, {Elsner}, {Leahy},
  {Weisskopf}, \& {Grindlay}}]{1982Natur.297..568S}
{Skinner}, G.~K., {Bedford}, D.~K., {Elsner}, R.~F., {et~al.} 1982, \nat, 297,
  568

\bibitem[{{Stairs} {et~al.}(2001){Stairs}, {Manchester}, {Lyne}, {Kaspi},
  {Camilo}, {Bell}, {D'Amico}, {Kramer}, {Crawford}, {Morris}, {Possenti},
  {McKay}, {Lumsden}, {Tacconi-Garman}, {Cannon}, {Hambly}, \&
  {Wood}}]{2001MNRAS.325..979S}
{Stairs}, I.~H., {Manchester}, R.~N., {Lyne}, A.~G., {et~al.} 2001, \mnras,
  325, 979

\bibitem[{{Tauris} \& {van den Heuvel}(2006)}]{2006csxs.book..623T}
{Tauris}, T.~M. \& {van den Heuvel}, E.~P.~J. 2006, in Compact stellar X-ray
  sources, ed. W.~H.~G. {Lewin} \& M.~{van der Klis} (Cambridge Astrophysics
  Series, No. 39, Cambridge, UK: Cambridge University Press), 623--665

\bibitem[{Vercellone(2017)}]{ctasurvey}
Vercellone, S. 2017, in {6th International
  Meeting on High-Energy Gamma-Ray Astronomy}, {AIP Conference Proceedings, 1792, 030001}

\bibitem[{{Walter} {et~al.}(2015){Walter}, {Lutovinov}, {Bozzo}, \&
  {Tsygankov}}]{2015A&ARv..23....2W}
{Walter}, R., {Lutovinov}, A.~A., {Bozzo}, E., \& {Tsygankov}, S.~S. 2015,
  \aapr, 23, 2

\end{thebibliography}
\listofobjects
\end{document}